\documentclass[%
nofootinbib,
 amsmath,amssymb,
 aps,
prd
]{revtex4-2}

\usepackage{graphicx}
\usepackage{dcolumn}
\usepackage{bm}
\usepackage{hyperref}

\usepackage[normalem]{ulem} 

\usepackage{xcolor,soul}

\DeclareMathOperator{\Div}{div}
\DeclareMathOperator{\Rot}{curl}
\DeclareMathOperator{\const}{const}
\renewcommand{\vec}{\mathbf}

\newcommand{\vp}{{\vec p}}

\renewcommand{\vr}{{\vec r}} 
\newcommand{\vrd}{{\pmb \rho}}

\newcommand{\vQ}{{\vec Q}}

\newcommand{\vVa}{{\vec V_\alpha}}   
\newcommand{\vVra}{{\vec V_{\alpha}}}   
   
\newcommand{\vVre}{{\vec V_{e}}}   
\newcommand{\vVrp}{{\vec V_{p}}}

\newcommand{\vj}{{\vec j}}       
\newcommand{\vA}{{\vec A}}       
\newcommand{\vB}{{\vec B}}

\newcommand{\vvg}{{\pmb v_\vp}}    
\newcommand{\vg}{{ v_\vp}}

\newcommand{\vv}{{\vec V_L}}     

\newcommand{\mN}{\mbox{$\mathcal N$}}

\begin{document}
\title{Non-purely transverse Magnus force in superconducting neutron stars}
\date{\today}
\author{Oleg A.~Goglichidze}
\email{goglichidze@gmail.com}
\author{Mikhail E.~Gusakov}
\affiliation{Ioffe institute, Politekhnicheskaya 26, 194021 St.~Petersburg, Russia}

\begin{abstract}

    The force acting on a proton vortex in extreme type-II superconducting neutron star matter is studied in the limit of vanishing temperature.
    A detailed analysis is presented on how momentum is transferred from length scales on the order of the London penetration depth to the vortex core. 
    To examine the momentum flux, expressions for proton and electron currents are derived for arbitrary distances from the vortex line. 
    It is shown that, in the regime of a large electron mean free path, the only force acting directly on the vortex core is the Magnus force. 
    Notably, the correction to the proton current near the vortex core generates a component of the Magnus force aligned with the incident current measured far from the vortex. 
    This contribution, responsible for longitudinal force, is usually overlooked in the literature. 
    The results obtained in this work for a relatively simple problem concerning the force on a vortex in cold matter
    of neutron stars  may also be relevant to other superconducting systems.
\end{abstract}

\maketitle


\section{Introduction}
\label{sec:intro}

    Neutron stars (NSs) are highly magnetized objects, exhibiting magnetic fields that span a wide range depending on their evolutionary stage -- from $\sim 10^{15}$~G in magnetars to $\sim 10^{12}$~G in ordinary radio pulsars, and further decreasing to $\sim 10^8$–$10^9$~G in millisecond pulsars.
    Developing a comprehensive theory to describe the evolution of these magnetic fields throughout the lifetime of NSs remains a major challenge for theorists. This problem is further complicated by the fact that, at sufficiently low temperatures, most of the NS core transitions into a type-II superconducting state, except in the deepest layers, where a type-I superconducting phase may occur \cite{GlampedakisAnderssonSamuelsson2011}.
    In type-II superconductors, the magnetic field is confined within proton vortices (flux tubes). 
    Understanding the forces acting on these vortices naturally emerges as a key subproblem in studying the evolution of NS magnetic fields.

    The forces on proton vortices have been the subject of numerous studies \cite{AlparLangerSauls1984,Jones1991b, Jones2006, Jones2009,AlfordSedrakian2010, GlampedakisAnderssonSamuelsson2011, GraberEtAl2015, BransgroveLevinBeloborodov2018,SourieChamel2020}, yet no  consensus has been reached regarding their exact form, even in the simplest case of $npe$-matter in  NS cores%
    %
    \footnote{By $npe$-matter, we refer to matter composed of neutrons ($n$), protons ($p$), and electrons ($e$).} 
    %
    at vanishing stellar temperature (when nucleon thermal excitations can be ignored).
    In this work, we examine the forces acting on vortices due to their motion relative to various matter components.
    For $npe$-matter at finite temperatures, up to five such components exist: superfluid  neutrons, superconducting protons, 
    neutron and proton normal components (thermal Bogoliubov excitations),
    and electrons. 
    Relative motion between vortices and these components can arise, for example, in the presence of a large-scale magnetic field that drives the system out of full thermodynamic equilibrium \cite{GusakovKantorOfengeim2020} or during stellar oscillations.
    At zero temperature, the number of matter components reduces to  three, since there are no thermal Bogoliubov excitations in the system.
    In the most general case, the force experienced by a vortex as it moves relative to different matter components can be written as
    \begin{equation}
        \label{eq:force_gen}
       \vec{F}_{\rm m \rightarrow v}  = \sum\limits_{\alpha} \left\{ - D_\alpha \vec{e}_z \times [\vec{e}_z\times (\vVa - \vv)] + D'_ \alpha \vec{e}_z\times (\vVa - \vv) \right\},
    \end{equation}   
    where the index $\alpha$ runs over all constituents, $\vVa$ is the velocity of the corresponding constituent at a large distance from the vortex line, $\vv$ is the velocity of the vortex line, $\vec{e}_z$ is the unit vector along the vortex, and $D_\alpha$ and $D'_\alpha$ are kinetic coefficients describing the interaction of different species with the vortex.
    Note that the velocities $\vVra$ are generally not independent,  which makes the definition of the coefficients $D_\alpha$ and $D'_\alpha$ somewhat ambiguous (see Sec.\ \ref{sec:force_calc}).

    Since the vortex introduces an axial vector into the system and breaks time-reversal symmetry, a perpendicular (or transverse) force component characterized by the coefficient $D'_\alpha$ may arise.
    This force may be associated with the vortex's response to the magnetic component of the Lorentz force acting on charged particles moving in the vortex magnetic field. Alternatively, it can be interpreted as the Magnus force exerted by superconducting protons. Additionally, a more subtle effect, known as the spectral flow force \cite{Jones2009}, can also contribute to the perpendicular force component.

    The existence of a transverse force has been debated: some studies suggest it vanishes 
    \cite{Jones1991b, Jones2006, Jones2009, BransgroveLevinBeloborodov2018}, while others argue it remains finite \cite{AlfordSedrakian2010, GlampedakisAnderssonSamuelsson2011, GraberEtAl2015}. 
    A review of this controversy can be found in Ref.~\cite{Gusakov2019}. In astrophysical literature, the force balance equation for a vortex is often derived through empirical reasoning and analogies with terrestrial superconductors. 
    However, the direct application of these results to matter in NS cores is problematic due to its unique properties, including very long mean free paths compared to other length scales in the system, a multi-component composition, the coexistence of proton and neutron condensates, strong nucleon-nucleon interactions, the absence of a crystalline lattice, etc. To address this problem, the definition of the force acting on a vortex must be formulated with sufficient rigor (an example of a related discussion in the context of vortices in Helium II can be found in Refs.\ \cite{Sonin1975, Hillel1981, HillelVinen1983}).

    In Ref.\ \cite{Gusakov2019}, the force on a proton vortex was calculated as the momentum flux through a cylindrical surface surrounding the vortex. The calculation was performed in the limit of  vanishing stellar temperature (see also Ref.\ \cite{SourieChamel2020}). It was shown that for a cylinder with a radius much larger than the London penetration depth, the only significant force on the vortex arises from electron scattering off the vortex magnetic field, allowing the determination of the coefficients $D_\alpha$ and $D'_\alpha$.
    On the other hand, assuming extreme type-II superconductivity, only a small fraction of the vortex’s magnetic flux is confined within its core. Consequently, in this region, the magnetic field has negligible impact, and the only force capable of transferring momentum directly into the vortex core is the Magnus force.%
    %
    \footnote{Strictly speaking, momentum can also be transferred through electron collisions with bound quasiparticles in the core. However, since the electron mean free path is vastly larger than the vortex core diameter (Sec.\ \ref{sec:hierarchy}), this effect is expected to be negligible.
    \label{scattering} 
    } 

    This situation poses a paradox.
    Indeed, the Magnus force is known to be strictly perpendicular to the superconducting proton current flowing through the vortex.
    That is, a naive application of the Magnus force formula using the incident (or transport) proton current yields a result identical to the perpendicular component of the electron-related force, obtained through integration over a cylinder of large radius and parametrized by the coefficients $D'_\alpha$.
    Thus, one can conclude that the transverse force is conserved: the perpendicular component of momentum crossing a large cylindrical surface is ultimately transferred to the vortex core by superconducting protons. However, there is also the parallel force component, defined by the coefficients $D_\alpha$. As the cylinder shrinks to the size of the vortex core, what happens to this component?
    In other words, what physical mechanism is responsible for transferring the parallel component of momentum to the vortex core?

    The primary goal of the present work is to address this question.
    To this end, we conduct a detailed study of electron scattering by the vortex's magnetic field. Specifically, we derive the electron distribution function at arbitrary distances from the vortex and analyze how electron scattering affects superconducting proton currents and, consequently, the Magnus force.

    The paper is organized as follows.
    In Sec.\ \ref{sec:hierarchy}, we discuss the characteristic length scales of the problem.
    In Sec.\ \ref{sec:basic_eqs}, considering the hierarchy of these length scales, we formulate the problem and introduce the set of equations modeling the system.
    Section \ref{sec:force_bal} is dedicated to the formal definition of the force acting on a vortex.
    In Sec.\ \ref{sec:inter}, we examine the disturbances induced in the surrounding matter by the vortex.
    Corrections to the electron distribution function (Sec.\ \ref{sec:scat_el}) and the proton  superconducting current (Sec.\ \ref{sec:protons}) are derived.
    Section \ref{sec:force_calc} focuses on the calculation of the force, employing two methods based on different definitions introduced in Sec.\ \ref{sec:force_bal}.
    Finally, in Sec.\ \ref{sec:summary}, we summarize the main results.
    In addition, the paper includes Appendices A–G, where several equations used in the text are derived, and some results that are not directly related to the main presentation are covered.

\section{Hierarchy of length-scales}
\label{sec:hierarchy}

    Let us start by discussing the length scales present in the problem. The vortices in Fermi liquids feature a core with a radius of the order of the coherence length, $\xi$. Inside the core, the order parameter drops to zero.%
    %
    \footnote{Note, however, that in terrestrial superfluids and superconductors, various unconventional vortex forms can exist, where the order parameter remains finite everywhere \cite{VolovikEltsovKrusius_inbook}.}
    %
    If the superfluid is charged, the vortex generates a magnetic field. 
    The size of the region occupied by the field is of the order of the London penetration depth $\lambda$. 
    These two parameters for the $npe$-mixture can be estimated as \cite{deGennes_book,Gusakov2019}
    \begin{equation}
	\label{eq:xi_def}
	\xi = \frac{\hbar p_{Fp}}{\pi \Delta^{(p)} m_p^*} \approx 28 \mbox{ fm} \left(\frac{n_p}{0.18\,  n_0}\right)^{1/3}
         \left(\frac{m_p}{m_p^*}\right) 
         \left(\frac{0.456 \mbox{ MeV}}{\Delta^{(p)}}\right),
    \end{equation}
    \begin{equation}
        \label{eq:lambda_def} 
	\lambda = \sqrt{\frac{m_p c^2}{4 \pi e^2 n_{p}} } \approx 42.4 \mbox{ fm} \left(\frac{0.18\,  n_0}{n_p} \right)^{1/2}.
    \end{equation}
    Here, $p_{Fp}$ is the proton  Fermi momentum, $m_p^*$ and $m_p$ 
    are, respectively,  the  effective and bare proton masses, $n_{p}$ is the proton number density, $\Delta^{(p)}$ is the proton energy gap, and $n_0 = 0.16$ fm$^{-3}$ is the nuclear saturation density. 
    Besides $\xi$ and $\lambda$ one can introduce the distance between neighboring vortices,
    \begin{equation}
	d_B = \sqrt{\frac{2 \Phi}{\sqrt{3} \bar{B}}} = 4.89\times 10^3\mbox{ fm} \sqrt{\frac{10^{12}}{ \bar{B}}},
    \end{equation}
    where $\Phi$ is the magnetic flux associated with the vortex and $ \bar{B}$ is the magnetic field induction averaged over an area containing a large number of the vortices. 
    For a typical NS $d_B \gg \lambda$. 
    This inequality allows us to reduce the problem of interaction of net flows with the collection of vortices to the calculation of the force acting on an isolated vortex. 
    In what follows, by large distances from the vortex line, we will mean the distances $\lambda \ll R \ll d_B$.

    Another important set of parameters is the particles wavelengths (divided by $2\pi$),
    \begin{equation}
	\frac{\hbar}{p_{F\alpha}} = 1.05 \mbox{ fm} \left(\frac{ 0.18\, n_0}{n_\alpha} \right)^{1/3},
    \end{equation}
    where $\alpha = n,\ p,\ e$.

    Finally, the system can be characterized by the electron mean free path $\ell$.%
    \footnote{ Here, $\ell$ represents the electron mean free path in the $npe$-mixture in the absence of vortices. }
    In a normal non-magnetized matter,
    it can be estimated as $\ell \sim 10^7\, T_8^{-1}$ fm \cite{ShterninOfengeim2022}, where $T_8 = T/ 10^8$~K.
    In the superconducting/superfluid matter it becomes even larger \cite{Shternin2018}.

\section{The model and basic equations}
\label{sec:prob_form}
\label{sec:basic_eqs}

    As it was argued in the previous section, we can consider an isolated vortex.
    Our primary aim is to examine the forces acting on the vortex, induced by the incident particle fluxes.
    These fluxes are assumed to be uniform and stationary, making the entire problem stationary.
    The vortex is assumed to be straight, enabling us to introduce cylindrical coordinates $(\rho, \phi, z)$ with the $z$-axis coinciding with the vortex axis and directed such that the integral \eqref{eq:circ} is positive, see below.

    We consider a mixture consisting of neutrons, protons, and electrons. 
    The modifications required to account for, e.g., muons are briefly discussed in Sec.\ \ref{sec:summary}.
    In the present paper, we focus on a special case where the neutrons and protons are strongly superfluid/superconducting, i.e., the presence of neutron and proton thermal excitations can be neglected.
    We also neglect the entrainment effect \cite{AndreevBashkin1976}
    throughout the main part of the paper. Its influence on the results  is discussed in Appendix \ref{sec:entr}.
    Another substantial simplification we adopt is assuming the vortex core is infinitely thin,
    meaning that we treat it as a line without considering its internal structure.
    In contrast, the London penetration depth is considered to be finite. 
    In other words, we are working under the condition that the coherence length is much smaller than the London penetration depth (the extreme type-II superconductivity).
    While this is not a very realistic assumption for the NS interiors (see previous section), it simplifies the analysis, allowing us to more clearly examine the physical processes associated with the forces under consideration.
    Such a simplification also prevents us from accounting for a force arising due to the scattering of electrons by proton quasiparticles bound in the vortex core \cite{KopninKravtsov1976b,SoninBook}.
    However, given the large mean free path compared to the other length-scales in the problem (see Sec.\ \ref{sec:hierarchy}), we expect this force to be a rather small correction \cite{Gusakov2019}.

    To describe the superconducting proton current, we introduce the condensate momentum \cite{AronovEtAl1981}
    \begin{equation}
	\label{eq:Q_def}
	\vQ_p = \frac{\hbar}{2} \nabla \chi_p - \frac{e}{c} \vec A 
    \end{equation}
    and the non-equilibrium proton chemical potential \cite{AronovEtAl1981, Gusakov2010, GG23}    
    \begin{equation}
	\label{eq:mu_def}
	\breve{\mu}_p = E_{Fp}-\frac{\hbar}{2} \frac{\partial \chi_p}{\partial t} - e \vec \varphi,
    \end{equation}
    where $\chi_p$ is the proton order parameter phase, $\varphi$ and $\vec A$ are the electrostatic  and vector potentials,  $e = |e|$ is the elementary charge, and by $E_{Fp}$ we denote the proton Fermi energy calculated in the absence of the vortex as well as any currents.
    The vortex manifests itself through a nonzero circulation of the phase gradient:
    \begin{equation}
	\label{eq:circ}
	\frac{\hbar}{2}\oint \nabla \chi_p d \vec{l} = m_p \varkappa,
    \end{equation}
    where the integral is calculated along an arbitrary contour  enclosing the vortex axis and
    \begin{equation}
        \label{eq:kappa_def}
	\varkappa = \frac{2 \pi  \hbar}{ 2  m_p}
    \end{equation}
    is the circulation quantum.

    The vector potential can be represented as $\vA = \vA_{v} + \delta \vA$, where $\vA_{v}$ is the vector potential generated by the vortex, while $\delta \vA$ is a small correction to $\vA_{v}$ arising due to the nonzero particle flows streaming through the vortex. 
    The vortex vector potential can be represented as \cite{Gusakov2019}
    \begin{equation}
	\label{eq:A}
	\vec{A}_v = \frac{c}{e}\frac{m_p \varkappa}{2 \pi \rho} \mathcal{P}(\rho) \vec{e}_\phi,
    \end{equation}
    where  the function $\mathcal{P}(\rho) $   has the following  properties:
    \begin{enumerate}
	\item  $\mathcal{P}(0) =0$;
	\item  $\mathcal{P}(\rho)/\rho \rightarrow 0 $ if $\rho\rightarrow 0$;
	\item $\mathcal{P}(\rho) \rightarrow 1  - a \left(\frac{\rho}{\lambda}\right)^{1/2} \exp \left( - \frac{\rho}{\lambda} \right)$ for $\rho\rightarrow 
        \infty$, where $a$ and  $\lambda$  are some constants. 
    \end{enumerate} 
    Here $\lambda$ can be interpreted as the London penetration depth. 
    To determine the exact form of the function $\mathcal{P}(\rho)$, one must solve the microscopic quantum problem in conjunction with Maxwell’s equations.
    However, for the case of extreme type-II superconductivity  ($\xi\ll\lambda$), an explicit solution can be obtained and is given by \cite{deGennes_book,Gusakov2019}
    \begin{equation}
        \label{eq:P_apr}
	\mathcal{P}(\rho) = 1 - \frac{\rho}{\lambda} K_1\left(\frac{\rho}{\lambda}\right),
    \end{equation}
    where $K_1$ is the modified Bessel function of the second kind, $K_m$, with index $m=1$.
    The vortex magnetic field, in turn, equals (see also Appendix \ref{sec:mag_field})
    \begin{equation}
	\label{eq:B_vortex}
	\vec{B}_v
	= \Rot \vec{A}_v = \frac{c}{e}\frac{m_p \varkappa}{2 \pi \rho} \mathcal{P}'(\rho)\vec{e}_z,
    \end{equation}
    where $\mathcal{P}'(\rho)$ denotes the derivative of the function $\mathcal{P}(\rho)$;
   in the limit $\xi\ll\lambda$, $\mathcal{P}'(\rho)$ equals
    \begin{equation}
	\label{eq:dP_apr}
	\mathcal{P}'(\rho)= \frac{\rho}{\lambda^2} K_0\left(\frac{\rho}{\lambda}\right).
    \end{equation}
    The total condensate momentum \eqref{eq:Q_def} can be represented as 
    \begin{equation}
	\label{eq:Q_mod_ch}
	\vQ_p
	= \vQ_{p0} + \vQ_{pv} + \delta \vQ_p.
    \end{equation}
    Here the first term, $\vQ_{p0} $, is the constant momentum describing the homogeneous proton flow measured far from the vortex; the second term,
    \begin{equation}
 	\label{eq:Q_pv_def}
 	\vQ_{pv} =  \frac{m_p \varkappa}{2 \pi \rho}\left[1 -  \mathcal{P}(\rho)\right] \vec{e}_\phi,
    \end{equation}
    is the vortex flow generated by the nonzero phase circulation; 
    and the third term, $\delta \vQ_p$, is a small correction, which should be determined self-consistently. 
    In the subsequent calculations, we will need  the curl of the condensate momentum, which equals
    \begin{align}
	\Rot\vQ_p  &= \frac{m_p \varkappa}{2 \pi \rho}\delta(\rho) \vec{e}_z- \frac{e}{c} \vec{B}
	\nonumber
	\\
	&=   -\frac{m_p \varkappa}{2 \pi \rho}  \left[\mathcal{P}'(\rho) - \delta(\rho) \right] \vec{e}_z - \frac{e}{c} \delta \vec{B}.
        \label{eq:rotQ_ch}
    \end{align}
    In Eq.\ \eqref{eq:rotQ_ch},
    we presented the total magnetic field as 
    \begin{equation}
        \label{eq:B_def}
        \vB = \vB_v + \delta \vB,
    \end{equation}
    where $\vB_v$ is the vortex magnetic field given by Eq.\ \eqref{eq:B_vortex} and 
    \begin{equation}
        \label{eq:dB_dA_rel}
        \delta \vec{B} = \Rot\delta \vec{A}
    \end{equation}
    is a small self-consistent correction.
    Some clarification is required regarding  Eq.\ (\ref{eq:rotQ_ch}).
    The first term on the right-hand side of the first line of this equation formally arises from taking the $\Rot$ of the superfluid phase gradient.
    It is well known from vector analysis that taking the $\Rot$ of a gradient of any well-defined, single-valued function yields identically zero.
    However, the phase $\chi_p$ and, consequently, the condensate momentum $\vQ_p$ are not defined (singular) at the vortex line. 
    This singularity arises due to the nonzero circulation of the phase gradient.
    Using identity~\eqref{eq:circ}, we define the formal construction $\Rot \nabla \chi_p$ at $\rho = 0$ by means of Stokes' theorem:
    \begin{equation}
	\frac{\hbar}{2} \int \Rot\nabla  \chi_p \vec{n} d{S} = \frac{\hbar}{2} \oint \nabla  \chi_pd\vec{l} =m_p \varkappa,
    \end{equation}
    where the left-hand integration is performed over an arbitrary (oriented) surface punctured by the vortex ($\vec{n}$ is the unit vector normal to the surface), while  the right-hand integration is taken over the oriented boundary of this surface.
    Thus, for a straight vortex, we have:  $\Rot \nabla \chi_p=2 \pi \delta^{(2)}({\vrd})\, \vec{e}_z$, where $\delta^{(2)}(\vrd) = \delta(x)\delta(y)$ is the two-dimensional delta function.
    To arrive at Eq.~\eqref{eq:rotQ_ch}, we applied the identity:
    \begin{equation}
        \label{eq:deltas_rel}
        \delta^{(2)}(\pmb \rho)=\frac{\delta(\rho)}{2 \pi \rho},
    \end{equation}
    which will also be useful in what follows.

    The proton current must satisfy the continuity equation
    \begin{equation}
	\label{eq:cont_eq_p}
	\Div  \vj_p = 0.
    \end{equation}
    If the contribution from proton thermal excitations is negligible, the proton current density takes the form:%
    %
    \footnote{For large $\vQ_p$, the Bogoliubov excitations exist even at vanishing temperature and the expression for the current can be more complicated \cite{gk13,Leinson2017,AllardChamel2021b,AllardChamel2023}. In our model, the region where this effect becomes substantial is included in the vortex core. See also the comment to Eq.~\eqref{eq:another_small_p}.
    }
    %
    \begin{equation}
        \label{eq:jp_def}
        \vj_p = \frac{n_p}{m_p} \vQ_p,
    \end{equation}
    where $n_p$ is the proton number density, and the proton condensate momentum is given by Eq.\ \eqref{eq:Q_mod_ch}.

    Let us consider, for a moment, a more general case where the vortex can move with a constant velocity $\vv$.
    The ``superfluid'' equation relating the quantities $\vQ_p$ and $\breve{\mu}_p$ in our system
    takes the following form \cite{Sonin1997}
    (see Appendix \ref{sec:p_sf_eq}):
    \begin{equation}
        \label{eq:sf_eq1}
        \frac{\partial \vQ_p}{\partial t} +  \nabla  \breve{\mu}_p  =   e   \vec E - \frac{\hbar}{2} \Rot \nabla \chi_p \times \vv,
    \end{equation}
    where 
    \begin{equation}
	\vec{E} =  - \nabla \varphi - \frac{1}{c} \frac{\partial \vec{A}}{\partial t}
    \label{El}
    \end{equation}
    is the electric field.
    Using Eqs.\ \eqref{eq:Q_def}, \eqref{eq:rotQ_ch} 
    \eqref{eq:deltas_rel},  and \eqref{eq:jp_def}
    and introducing the time-dependent position of the vortex $\vrd_L(t)$ (cf.\ Appendix \ref{sec:p_sf_eq}), we can  rewrite the same equation in a more familiar, Euler-like fashion \cite{Sonin1997}: 
    \begin{equation}
        \label{eq:sf_eq2}
        \frac{\partial \vQ_p}{\partial t} + \frac{1}{m_p} ( \vQ_p \nabla ) \vQ_p + \nabla \left( \breve{\mu}_p - \frac{\vQ_p^2}{2 m_p}\right)   
        = \vec{f}_{\rm Lp} - \vec{f}_{\rm M},
    \end{equation}
    where 
    \begin{equation}
        \label{eq:Lorenz_def}
        \vec{f}_{\rm Lp} =  e \left(  \vec E +  \frac{1}{ n_p c} \vj_p \times \vB \right)
    \end{equation}
    and
    \begin{equation}
        \label{eq:Magnus_per_p}
        \vec{f}_{\rm M} = - {m_p \varkappa} \, \delta^{(2)}(\pmb \rho - \pmb \rho_L(t)) \ \vec{e}_z \times\left( \frac{\vj_p}{n_p} - \vv \right).
    \end{equation}
    In Eq.\ \eqref{eq:sf_eq2}, the combination $\mu_{p,\rm phen} =  \breve{\mu}_p - {\vQ_p^2}/{2 m_p}$  is usually defined as the chemical potential in the phenomenological hydrodynamics of superfluid helium-4 \cite{Khalatnikov_book}.
    It can be interpreted as the chemical potential measured in the frame moving with the condensate. 
    The first term on the right-hand side of Eq.\ \eqref{eq:sf_eq2}, given by Eq.\ \eqref{eq:Lorenz_def}, represents the standard Lorentz force acting on the superconducting protons. 
    As for the second term, given by Eq.~\eqref{eq:Magnus_per_p}, one can recognize the Magnus force (per one proton), exerted on the infinitely thin vortex core.
    This term appears in Eq.\ \eqref{eq:sf_eq2} with a negative sign because the protons experience the {\it reaction} force associated with $\vec{f}_{\rm M}$.
    The physical conditions that give rise to this force are discussed in Sec.~\ref{sec:force_bal}.
    The ``superfluid'' equation in the form of Eq.~\eqref{eq:sf_eq2}  serves as a convenient basis for deriving the momentum conservation equation (see Appendix~\ref{sec:mom_cons}).

    Returning to our stationary problem where the vortex is at rest along the $z$-axis, Eq.\ \eqref{eq:sf_eq1} simplifies to
    \begin{equation}
	\label{eq:sf_eq}
	\nabla  \breve{\mu}_p  = - e \nabla  \varphi =   e   \vec E.
    \end{equation}
    If there are no incident fluxes, the electric field is zero \cite{Gusakov2019} and, therefore, the function $\breve{\mu}_p$ remains unperturbed.
    Thus, accounting for the definition \eqref{eq:mu_def}, we can represent $\breve{\mu}_p$ as
    \begin{equation}
        \label{eq:mup_repr}
        \breve{\mu}_p = E_{Fp} + \delta \breve{\mu}_p,
    \end{equation}
    where $\delta \breve{\mu}_p$ is a small correction caused by the interaction of the vortex with the incident flux.
    This small correction should be determined self-consistently  together with the electric field.

    Turning to the neutrons, we can write down the following equations, which are valid in the proton vortex rest frame:
    \begin{equation}
        \label{eq:neutron_sf_eq}
         \nabla  \breve{\mu}_n  =   0,
    \end{equation}  
    \begin{equation}
        \Div \vj_n = 0,
    \end{equation}
    where $\breve{\mu}_n$ is the neutron non-equilibrium chemical potential, 
    \begin{equation}
        \label{eq:jn_def}
        \vj_n = \frac{n_n}{m_n} \vQ_n
    \end{equation}
    is the neutron current density, $n_n$ is the neutron particle density, and $\vQ_n$ is the neutron condensate momentum.
    All neutron-related quantities are defined similarly to their proton counterparts, but with the electric charge set to zero ($e=0$).
    In the absence of the entrainment effect, the neutron condensate momentum reduces to $\vQ_n = \vQ_{n0} = \const$.

    Let us finally move to the electrons.
    Since the typical electron wavelength is small compared to other length scales (see Sec.\ \ref{sec:hierarchy}), we can treat electrons as a classical ideal relativistic gas.
    Therefore, the electron distribution function satisfies the  Boltzmann-Vlasov equation:
    \begin{equation}
	\label{eq:kin_eq_e}
				\vvg \frac{\partial \mN_{\vp}^{ (e)}}{\partial \vr} 
				- e \left( \vec{E} +\frac{1}{c} \vvg \times \vec{B} \right) \frac{\partial \mN_\vp^{ (e)}}{\partial \vp} =  I^{ (e)},
    \end{equation}
    where $\vvg = \vp c^2 / \varepsilon_\vp$ and $ \varepsilon_\vp = \sqrt{m_e^2 c^4 + p^2 c^2}$.
    As argued above, for our particular problem, collisions are unimportant and, hence, the collision integral $I^{(e)}$ in Eq.\ \eqref{eq:kin_eq_e} can be omitted.

    Considering the superconducting mixture, we expect that far from the vortex, the total electric current vanishes 
    due to the Meissner effect \cite{Jones1991,Jones2006,GlampedakisAnderssonSamuelsson2011,GusakovDommes2016}.
    This screening condition leads to the relationship:
    \begin{equation}
        \label{eq:screen_cond_npe}
        \vj_{p0} = \vj_{e0},
    \end{equation}
    where $\vj_{p0}$ and $\vj_{e0}$ are the proton and electron particle current densities in the absence of the vortex. These quantities are defined, respectively, by Eqs.\ \eqref{eq:je0_def} and \eqref{eq:jp0_def} in what follows.

\section{Two expressions for the force on the vortex}
\label{sec:force_bal}

    \begin{figure}
	   \includegraphics[width=0.3\textwidth,trim= 5.5cm 3.3cm 5.0cm 3.4cm, clip = true]{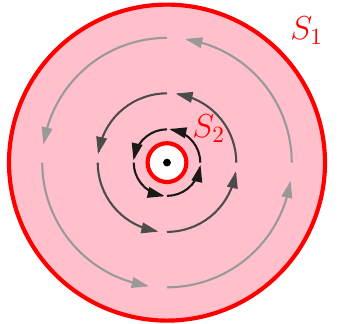}
       \hspace{2.5cm}
	\includegraphics[width=0.3\textwidth,trim= 5.5cm 3.3cm 5.0cm 3.4cm, clip = true]{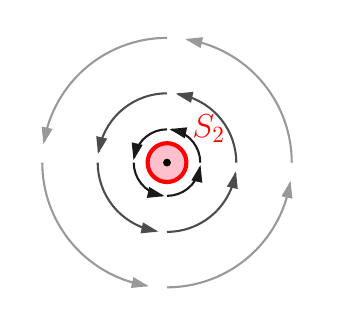}      
	\caption{
         \label{fig:volumes}
                    Sketch of the vortex and integration volumes.
                    The vortex flow $\vQ_{pv}$ is schematically shown by circular arrows; 
                    the infinitely thin vortex core is indicated by a dot at the center of the vortex.
                    The cylindrical surfaces $S_1$ and $S_2$ are depicted as red circles, and the integration volumes are shaded in pink.
            }
    \end{figure}

    In this section, we analyze the forces acting in our system and determine the force applied to the vortex.
    We start with the momentum conservation equation, 
    which takes the following form (see Appendix~\ref{sec:mom_cons}):
    \begin{align} 
        \frac{\partial \mathfrak{P}^i}{\partial t}
	+ \frac{\partial \Pi^{ik}}{\partial r^k}
	= - n_p {f}_{\rm M}^i, 
        \label{eq:mom_eq_ap} 
    \end{align}
    where the total momentum density $\mathfrak{P}$ and the total stress tensor $\Pi^{ik}$  consist of contributions from superfluid and superconducting $npe$-matter as well as the electromagnetic field.
    The exact expressions for these quantities are given by Eqs.\ \eqref{eq:total_mom} and \eqref{eq:total_stress}, respectively.
    Note that the Lorentz forces acting on electrons and protons are internal to our system, which includes the electromagnetic field.
    Thus, they can be expressed in terms of the derivatives of the electric and magnetic fields and incorporated into the left-hand side of Eq.\ \eqref{eq:mom_eq_ap}.

    Looking at Eq.\ \eqref{eq:mom_eq_ap}, we see that the momentum in our system is actually not conserved:
    the right-hand side of this equation contains an additional force, $- n_p\vec{f}_{\rm M}$, acting on the system.
    As we found in Sec.\ \ref{sec:prob_form}, this force is localized on the vortex axis and is equal 
    to the 
    Magnus force taken with the opposite sign.
    We will return to discussing this force in what follows.

    Turning to the question of the forces acting on the vortex, it is most convenient to perform our analysis in the reference frame in which the vortex is at rest.
    For the stationary problem, we can apply Eq.\ \eqref{eq:mom_eq_ap} with the time derivative set to zero.
    We also assume that the vortex axis coincides with the $z$-axis [i.e., we set $\vrd_L=0$ and $\vv=0$ in Eq.\ \eqref{eq:Magnus_per_p}].

    Let us choose two unit-length cylinders coaxial with the vortex, denoted as $S_1$ and $S_2$ (see the left panel in Fig.~\ref{fig:volumes}).
    Integrating the stationary equation \eqref{eq:mom_eq_ap} over the volume bounded by $S_1$ and $S_2$ and applying Gauss's theorem, we obtain:
    \begin{equation}
        \label{eq:FSn_rel}
        \vec{F}_{R_1} = \vec{F}_{R_2},
    \end{equation}
    where
    \begin{equation}
        \label{eq:FSn_def}
        {F}_{R_n}^i = - \oint\limits_{\rho = R_n} \Pi^{ik} n_k dS
    \end{equation}
    is the  momentum flux entering the cylinder $S_n$ ($n = 1,\ 2$).
    In Eq.\ \eqref{eq:FSn_def}, the integration is performed over the lateral surface of the cylinder.
    Here, $R_n$ is the radius of the cylinder $S_n$, and $\vec{n} = \vec{e}_\rho$ is the outward-pointing unit normal vector at the cylinder's surface.
    In deriving Eqs.\ \eqref{eq:FSn_rel} and \eqref{eq:FSn_def}, we assumed that the stress tensor $\Pi^{ik}$ does not depend on the $z$-coordinate. 
    Therefore, integration over the top and bottom surfaces of the cylinders can be omitted.
    Since the radii $R_1$ and $R_2$ in Eq.\ \eqref{eq:FSn_rel} are chosen arbitrarily, we conclude that the integral \eqref{eq:FSn_def} is independent of $R_n$.

    The momentum flux entering the cylinder $S_n$ can be interpreted as the force exerted by the matter outside $S_n$ on the matter enclosed within it.
    In our model, the vortex possesses an infinitely thin core.
    Thus, in the limit $R_2 \rightarrow 0$, the integral $\vec{F}_{R_2}$ can be understood as the force applied directly to the vortex core.%
    %
    \footnote{ In reality, the radius $R_2$ cannot be chosen smaller than the physical size of the  vortex core  $\sim \xi$.}
    %
    On the other hand, since the problem is stationary, momentum cannot accumulate inside $S_2$.
    This means that another (external) force must be applied to the vortex core to drain momentum from it.
    Thus, we have the following force balance equation for the core \cite{Gusakov2019}:
    \begin{equation}
        \label{eq:force_bal_core}
        \vec{F}_{R_2} + \vec{F}_{\rm ext} = 0,
    \end{equation}    
    where $\vec{F}_{\rm ext}$ is the draining force that is external to our system. 
    We can extract more information 
    about the force $\vec{F}_{\rm ext}$   if we integrate Eq.\ \eqref{eq:mom_eq_ap} over the volume of the cylinder $S_2$ (see the right panel in Fig.\ \ref{fig:volumes}).
    Again, using Gauss's theorem, we obtain:
    \begin{equation}
        \label{eq:force_bal_core2}
        \vec{F}_{R_2}   - \vec{F}_{\rm M} = 0,
    \end{equation}
    where 
    \begin{equation}
        \label{eq:F_M_def}
        \vec{F}_{\rm M} 
        =  \int\limits_{S_2} d^2 \rho \, n_p \vec{f}_{\rm M}.
    \end{equation}
    Here, in Eq.~\eqref{eq:F_M_def}, the volume integral is replaced with a two-dimensional integral multiplied by the unit cylinder length.
    Comparing Eqs.~\eqref{eq:force_bal_core} and \eqref{eq:force_bal_core2}, we conclude that 
    \begin{equation}
        \label{eq:force_bal_core3}
        \vec{F}_{\rm ext} = - \vec{F}_{\rm M}.
    \end{equation}
    Thus, we find that the external force $\vec{F}_{\rm ext}$ applied to the vortex core
    is equal to the Magnus force taken with the opposite sign. 
    At first glance, this result may seem  puzzling.
    Indeed, the Magnus force in Eq.\ \eqref{eq:mom_eq_ap} originates from one of the equations describing our system -- the Euler-like equation \eqref{eq:sf_eq2} for superconducting protons.
    However, how does this force ``know'' that it must be equal to some external force $\vec{F}_{\rm ext}$?
    To answer this question, let us note that the Magnus force depends on the vortex velocity $\vv$, which is not determined by the dynamic equations and  is a {\it predefined} external parameter for our system.
    Accordingly, Eq.~\eqref{eq:force_bal_core3} relates two external parameters, $\vec{F}_{\rm ext}$ and $\vv$, effectively stating that the vortex cannot move with an arbitrary velocity $\vv$ relative to the surrounding matter unless an external force, $\vec{F}_{\rm ext}$, is applied to the vortex core.
    One can say that the external force $\vec{F}_{\rm ext}$ is parameterized by the velocity $\vv$ in such a way that it satisfies Eq.~\eqref{eq:force_bal_core3}.

    As follows from Eqs.\ \eqref{eq:Magnus_per_p} and \eqref{eq:force_bal_core3}, in the absence of an external force ($\vec{F}_{\rm ext} = -\vec{F}_{\rm M} = 0$), the vortex is entrained by the proton superflow, $\vv = \vj_p /n_p$,%
    %
    \footnote{ Strictly speaking, the proton current diverges at the vortex line [see Eqs.~\eqref{eq:jp_def} and 
\eqref{eq:Q_pv_def}]. 
    As shown in Sec.~\ref{sec:Method_II}, one should substitute the regular part of $\vj_p$.
    }
    %
    which is a well-known result \cite{donnelly05, SoninBook}. 
    Only in this case can our system be considered truly isolated, meaning that the source term $-n_p \vec{f}_{\rm M}$ in Eq.~\eqref{eq:mom_eq_ap} vanishes and momentum is strictly conserved.

    Returning to Eqs.\ \eqref{eq:FSn_rel}--\eqref{eq:force_bal_core2}, we note that by specifying the external force $\vec{F}_{\rm ext}$, we unambiguously determine the force $\vec{F}_{R_2}$ (and hence all other forces $\vec{F}_{R}=\vec{F}_{R_2}$, obtained by integrating the momentum flux over any cylinder of arbitrary radius $R$).
    On the other hand, according to Eq.\ \eqref{eq:force_bal_core2}, the force exerted on the vortex core by the surrounding matter is the Magnus force, applied by superconducting protons.
    Thus, we can conclude that regardless of the specific carrier of momentum at a distant surface $S_1$, in the immediate vicinity of the vortex core the {\it entire momentum flux is absorbed by the superconducting protons, which transfer it to the core}.%
    %
    \footnote{
    This conclusion is a direct consequence of our 
    approximations and is not general.
    For example, in this work, we (justifiably) ignore electron scattering off localized Bogoliubov quasiparticles in the vortex core (see footnote~\ref{scattering}).
    If such scattering were important, it would lead to the direct transfer of some fraction of electron momentum to the vortex core.
    This effect could be accounted for by including a collision integral in the electron kinetic equation, describing electron scattering off localized quasiparticles.
    As a result, in the right-hand side of the momentum conservation law \eqref{eq:mom_eq_ap}, an additional term responsible for such scattering would appear alongside the term $-n_p {\pmb {\rm f}}_M$.
    Consequently, the vortex would experience not only the Magnus force from the surrounding matter but also an additional force exerted by electrons.
    }
    %

    Summarizing the results of this section, the problem can be formulated as follows. 
    A proton vortex streamlined by the surrounding matter is held in place by an external force.%
    %
    \footnote{In NSs, such a force can be provided by buoyancy or tension effects \cite{Gusakov2019,dg17},
    as well as by pinning to neutron vortices.}
    %
   Our aim is to study the force exerted on the vortex by the incident particle flux.
    This force is defined as    
    \begin{equation}
	\label{eq:F_mv_def}
	{F}_{\rm m \rightarrow v}^i = - \oint\limits_{\rho = R} \Pi^{ik} n_k dS ,
    \end{equation}
    where, due to property~\eqref{eq:FSn_rel}, the radius $R$ can be chosen arbitrarily for convenience.
    This force can be calculated using two methods. 
    Method I directly employs Eq.~\eqref{eq:F_mv_def}.
    Method II, utilizing equality~\eqref{eq:force_bal_core2}, provides an alternative expression for the same force:
    \begin{equation}
	\label{eq:F_mv_Meth2}
	\vec{F}_{\rm m \rightarrow v} = -  {m_p \varkappa} \ \int d^2 \rho \,  \frac{\delta(\rho)}{2\pi \rho} \left(\vec{e}_z\times {\vec{j}_p} \right).
    \end{equation}
    Here, we substituted Eq.~\eqref{eq:Magnus_per_p} and used identity \eqref{eq:deltas_rel}.
    In this paper, we apply both methods to verify that they yield equivalent results.

\section{Interaction of the surrounding matter with the vortex}
\label{sec:inter}

    Before calculating the force, we must first determine the distribution function of the electrons in the presence of a vortex, as well as the proton current density, $\vj_p$.
    In this section, we derive these quantities with the required level of accuracy.

\subsection{Electrons}
\label{sec:scat_el}

    Let us start with the electrons.
    The problem of electron scattering off a proton vortex was studied in Ref.~\cite{Gusakov2019}, where an asymptotic expression for the electron distribution function was obtained far from the vortex.
    In this section, we generalize these results by finding the distribution function at any distance from the vortex.
    Electrons scatter off the vortex via interaction with its magnetic field.
    Besides the magnetic field, electrons can also experience the electric field induced by perturbations in the particle densities of the charged constituents.
    The electron distribution function can be found as a solution to Eq.\ \eqref{eq:kin_eq_e} with a zero collision integral $I_e$.

    We assume that, at an infinite distance from the vortex axis, the electron distribution function reduces to the standard Fermi distribution.
    At finite distances, we shall look for the distribution function in the following form:
    \begin{equation}
	\label{eq:Ne_ans}
	   \mN_{\vp}^{ (e)} = f_F\left( \varepsilon_\vp^{(e)} - E_{Fe} + \frac{m_e \vVre^2}{2} - e \varphi - \vp  \vVre +  g_e \right),
    \end{equation}
    where 
    \begin{equation}
	f_F(x) = \frac{1}{e^{x/T} +1 }
    \end{equation}
    is the Fermi distribution function,
    $E_{Fe}$ is the electron (unperturbed) Fermi energy,
    $\varphi$ is the electrostatic potential,
    $\vVre$ is the velocity of the electron incident 
    flux,
    and $g_e$ is a function of $\vr$ and $\vp$ that needs to be determined.
    Without any loss of generality, we can assume that $\vVre \perp \vec{e}_z$, as the flow parallel to the vortex does not contribute to the force.
    Substituting the ansatz \eqref{eq:Ne_ans} into Eq.\ \eqref{eq:kin_eq_e}, we obtain an equation for the function $g_e$:
    \begin{equation}
	\label{eq:dpe_eq_tmp1}
	\left(\vvg \frac{\partial }{\partial \vr} \right) g_e
	- e \left[\left( \vec{E} +\frac{1}{c} \vvg \times \vec{B} \right) \frac{\partial }{\partial \vp} \right] g_e
	= - \frac{e}{c} \vVre  \left( { \vvg}\times \vec{B}\right) - e \vec{E} \vVre.
    \end{equation}
    We assume that the velocity of incident electron flux is small, namely 
    \begin{equation}
	\label{eq:e_ineq}
	\frac{V_{e}}{ v_{Fe} } \ll 1,
    \end{equation}
    where $v_{Fe}$ is the electron Fermi velocity.
    The same is also assumed for the proton incident flux, see Eq.\ \eqref{eq:p_ineq} below.
    In view of these assumptions, we maintain linear accuracy in incident fluxes throughout the paper.
    After neglecting the higher order terms, Eq.\ \eqref{eq:dpe_eq_tmp1} becomes 
    \begin{equation}
	\label{eq:dpe_eq_tmp2}
	\left(\vp \frac{\partial }{\partial \vr} \right)  g_e
	- \frac{e}{c} \left[ (\vp \times \vec{B}_v)  \frac{\partial }{\partial \vp} \right]  g_e
	= - \frac{e}{c} \vVre  \left(\vp\times \vec{B}_v \right),
    \end{equation}
    where we took into account that the electric field $\vec{E}$ as well as the self-consistent correction to the magnetic field $\delta \vec{B}$, are at least linear in the incident fluxes.

    Besides the ratios \eqref{eq:e_ineq} and \eqref{eq:p_ineq},
    Eq.\ \eqref{eq:dpe_eq_tmp2} contains another small parameter 
    \begin{equation}
        \tilde{\epsilon} =  \frac{\hbar}{p_{Fe} \lambda} \sim \frac{e}{c} B_v \frac{\lambda}{p_{Fe}} \ll 1,
    \end{equation}
    where $B_v$ is estimated using Eq.\ \eqref{eq:B_vortex}.%
    %
    \footnote{The smallness of the parameter $\tilde{\epsilon}$ means that the electrons pass through the vortex along  trajectories that are almost rectilinear.}
    %
    For our problem, however, it is more convenient to introduce an alternative parameter
    \begin{equation}
	\label{eq:eps_p_def}
	\epsilon =     \frac{m_p\Delta^{(p)}}{ p_{Fp}^2} = m_p \varkappa \frac{1}{\pi^2 p_{Fp} \xi}
    \end{equation}
    instead of $\tilde{\epsilon}$.
    This choice may not seem very natural for the problem of electron scattering.
    However, we will treat $\epsilon$ as a universal small parameter for both electrons and protons. 
    As we will see,  $\epsilon$ emerges more naturally in the case of protons.
    Using Eq.\ \eqref{eq:xi_def} and 
    bearing in mind
    that $p_{Fe} = p_{Fp}$, we establish a relationship between these parameters:
    \begin{equation}
        \tilde{\epsilon} = \pi \frac{\xi}{\lambda} \epsilon.
    \end{equation}
    Since $\xi < \lambda$, the smallness of $\epsilon$ ensures that $\tilde{\epsilon}$ is also small.
    Thus, we can expand the function $g_e$ in  powers of $\epsilon$: $g_e = g_{e1} + g_{e2} + ...$\,. 
    The first two terms in this expansion satisfy the following equations:
    \begin{align}
	\label{eq:dpe1_eq}
	&\left(\vp \frac{\partial }{\partial \vr} \right)  g_{e1}
	= - \frac{m_p \varkappa}{2 \pi \rho} \mathcal{P}'(\rho) \,   \vp \left( \vec{e}_z \times \vVre \right)   ,
	\\
	\label{eq:dpe2_eq}
	&\left(\vp \frac{\partial }{\partial \vr} \right)  g_{e2}
	=  \frac{m_p \varkappa}{2 \pi \rho}\mathcal{P}'(\rho)  \left[ \left(\vp \times \vec{e}_z\right)  \frac{\partial }{\partial \vp} \right] g_{e1},
    \end{align}
    where we made use of Eq.\ \eqref{eq:B_vortex}.
    It is easy to see from these equations that the expansion in $\epsilon$ is formally equivalent to an expansion in the dimensional parameter $m_p\varkappa$.
    We seek a solution that does not depend on $z$-coordinate and vanishes at infinity.
    The exact solution is provided in Appendix \ref{sec:sol_to_e}. 
    At distances much larger than the London penetration depth $\lambda$, 
    it reduces to
    (see Appendix \ref{sec:sol_to_e} for details; cf.\ the results of Ref.\ \cite{Gusakov2019}):
    \begin{equation}
	\label{eq:dpe_ld}
	g_{e} =  \left\{  - \left[\vp \left( \vec{e}_z \times \vVre \right) \right] \, \sigma_{\perp}^{(e)} +\left( \vp_\perp \vVre \right) \, \sigma_{||}^{(e)} \right\} \frac{\delta(\phi-\phi_p)}{\rho},
    \end{equation}
    where we introduced, respectively, the transverse and transport cross-sections \cite{Sonin1975,AronovEtAl1981,Gusakov2019}:
    \begin{align}
	\label{eq:sigma_e_perp}
	\sigma_{\perp}^{(e)}& = \frac{m_p \varkappa }{p_\perp},
	\\
	\label{eq:sigma_e_par}
	\sigma_{||}^{(e)}  &=  \frac{(m_p \varkappa)^2 }{p_\perp^2}  \frac{1}{8\pi^2}\int\limits_{-\infty}^{\infty} dx \left( \int \limits_{-\infty}^{\infty}      dy\frac{\mathcal{P}'(\sqrt{x^2+y^2})}{\sqrt{x^2+y^2}}   \right)^2.
    \end{align}
    In these expressions, $\vp_\perp$ is the component of the vector $\vp$ perpendicular to $\vec{e}_z$, 
    $\varphi_p$ is its azimuthal angle (see Fig.~\ref{fig:coords_2d}), and $\delta$ is the Dirac delta function. 

    For the subsequent calculations, we will also require the electron particle current density, which can be calculated as 
    \begin{equation}
	\vj_e = \sum_{\vp \sigma} \vvg  \mN_{\vp}^{(e)}.
    \end{equation}
    Substituting the ansatz \eqref{eq:Ne_ans}, we obtain, to linear accuracy in $\vVre$,
    \begin{equation}
	\label{eq:je_exp}
	\vj_e = \vj_{e0}  + \delta \vj_e,
    \end{equation}
    where
    \begin{equation}
	\label{eq:je0_def}
	\vj_{e0} = n_{e0} \vVre
    \end{equation}
    is the electron current density at large distances (also referred to as the electron transport current),
    $n_{e0} = \sum_{\vp \sigma}   \mN_{\vp,0}^{(e)}$ is the unperturbed electron particle density, 
    $ \mN_{\vp,0}^{(e)} = f_F\left( \varepsilon_\vp^{(e)} - E_{Fe}\right)$, and 
    \begin{equation}
	\label{eq:dje_def}
	\delta \vj_e = \sum_{\vp \sigma} \vvg \, g_e \frac{\partial \mN_{\vp,0}^{(e)} }{\partial  \varepsilon_\vp^{(e)}}
    \end{equation}
    is the correction to the  current density due to electron scattering off the vortex. 
    In calculating the current density correction, we restrict ourselves to linear accuracy in the small parameter $\epsilon$. 
    This means that we can substitute the expression for $g_{e1}$ given by Eq.\ \eqref{eq:mn1_el_ap} into Eq.\ \eqref{eq:dje_def}. 
    Approximating the derivative of the electron distribution function as
    \begin{equation}
        \label{eq:dN_approx}
        \frac{\partial \mN_{\vp,0}^{(e)} }{\partial  \varepsilon_\vp^{(e)}} \approx - \delta(\varepsilon_\vp^{(e)} - E_{Fe}),
    \end{equation}
    we obtain 
    \begin{equation}
	\label{eq:dvj_e}
	\delta j_e^i= - J_e^{ik} \, (\vec{e}_z\times \vj_{e0})^k,
    \end{equation}
    where Eq.~\eqref{eq:je0_def} was taken into account 
    and the following tensor was introduced:
    \begin{equation}
	\label{eq:Je_tmp}
	J_e^{ik} = - m_p \varkappa  \, \frac{3}{4 \pi p_{Fe}}    \int d\Omega_{\vp} {e}_{\vp}^i {e}_{\vp_\perp}^k
	   \int \limits_{-\infty}^{ \rho \cos(\phi_p - \phi)} dy_1 \frac{{\mathcal{P}}'\left(\sqrt{\rho^2 \sin^2(\phi_p - \phi)+y_1^2}\right)}{2 \pi \sqrt{\rho^2 \sin^2(\phi_p - \phi)+y_1^2}} .
    \end{equation}  
    In the last expression, $\vec{e}_\vp = \vp/p$, $\vec{e}_{\vp_\perp} = \vp_\perp/p_\perp$, 
    angles $\phi$ and $\phi_p$ are defined as shown in Fig.~\ref{fig:coords_2d},
    and the integration is performed over the solid angle of the vector $\vp$.
    The tensor \eqref{eq:Je_tmp} can be rewritten in a more symmetrical form.
    To do this, we present it as: $J_e^{ik} = (1/2) J_e^{ik} + (1/2)J_e^{ik}$. 
    In the second term, we perform the following substitutions sequentially: first, we replace $\phi_p \rightarrow \phi_p + \pi$ during the integration over $\Omega_p$, and then we replace $y_1 \rightarrow - y_1$ during the integration over $y_1$. After applying these transformations, the two terms can be combined, yielding:
    \begin{equation}
	\label{eq:Je_def}
	J_e^{ik} = 
        -   m_p \varkappa \, \frac{3}{8 \pi p_{Fe}}   \int d\Omega_{\vp} {e}_{\vp}^i {e}_{\vp_\perp}^k
	   \int \limits_{-\infty}^{ \infty} dy_1 \frac{{\mathcal{P}}'\left(\sqrt{\rho^2 \sin^2(\phi_p - \phi)+y_1^2}\right)}{ 2 \pi \sqrt{\rho^2 \sin^2(\phi_p - \phi)+y_1^2}} .
    \end{equation}
    Expression~\eqref{eq:dvj_e} can be further simplified.
    To do that, we express
    the momentum-directed unity vectors as
    \begin{align}
      & \vec{e}_\vp =  \sin\theta_p \cos(\phi_p - \phi) \vec{e}_\rho + \sin\theta_p \sin(\phi_p - \phi) \vec{e}_\phi + \cos \theta_p \vec{e}_z, 
       \\
      & \vec{e}_{\vp\perp} =   \cos(\phi_p - \phi) \vec{e}_\rho +  \sin(\phi_p - \phi) \vec{e}_\phi,
    \end{align}
    where $\theta_p$ is the polar angle of the vector $\vp$.
    Substituting these representations into Eq.\ \eqref{eq:Je_def} and replacing $\phi_p \rightarrow \phi_p + \phi$ during the integration over the momentum azimuthal angle, we obtain:
    \begin{equation}
        \label{eq:dje_alt_repr}        
        \delta \vj_e=  - J_\rho (\rho) \ ( \vj_{e0} \, \vec{e}_\phi)\,  \vec{e}_\rho +  J_\phi (\rho) \ (\vj_{e0} \, \vec{e}_\rho) \,  \vec{e}_\phi,
    \end{equation}
    where
    \begin{equation}
        \label{eq:J_rho}
        J_\rho (\rho) =  m_p \varkappa \frac{3}{16 p_{Fe}}  \int\limits_0^{2\pi} d\phi_p\cos^2\phi_p  \int \limits_{-\infty}^{ \infty} dy_1 \frac{{\mathcal{P}}'\left(\sqrt{\rho^2 \sin^2\phi_p+y_1^2}\right)}{2 \pi\sqrt{\rho^2 \sin^2\phi_p+y_1^2}},
    \end{equation}
    \begin{equation}
         \label{eq:J_phi}
        J_\phi (\rho) =   m_p \varkappa \frac{3}{16 p_{Fe}}  \int\limits_0^{2\pi} d\phi_p\sin^2\phi_p  \int \limits_{-\infty}^{ \infty} dy_1 \frac{{\mathcal{P}}'\left(\sqrt{\rho^2 \sin^2\phi_p+y_1^2}\right)}{2 \pi \sqrt{\rho^2 \sin^2\phi_p+y_1^2}}.
    \end{equation}

    In the vicinity of the vortex core, the functions $J_\rho (\rho)$ and $J_\phi (\rho)$ can be approximated by their values at the vortex axis.
    For $\rho=0$, the integral over $y_1$ in Eqs.~\eqref{eq:J_rho} and \eqref{eq:J_phi} evaluates to $1/2\lambda$ for the the function $\mathcal{P}'$ given by Eq.~\eqref{eq:dP_apr}.
    Taking this into account, it is easy to  verify that the electron current density correction 
    is approximately given by
    \begin{equation}
        \label{eq:dje_ld}
        \delta \vj_e(\vrd) \approx \delta \vj_e(0) =   \frac{3 m_p \varkappa}{16 p_{Fe}} \frac{\pi}{2 \lambda} \, \vec{e}_z \times \vj_{e0}.
    \end{equation}
    Thus, as $\rho \rightarrow 0$, the electron current correction remains finite and is directed perpendicular to the incident current.
 
\begin{figure}
	\includegraphics[width=0.4\textwidth,trim= 0.8cm 1.5cm 2.7cm 1.5cm, clip = true]{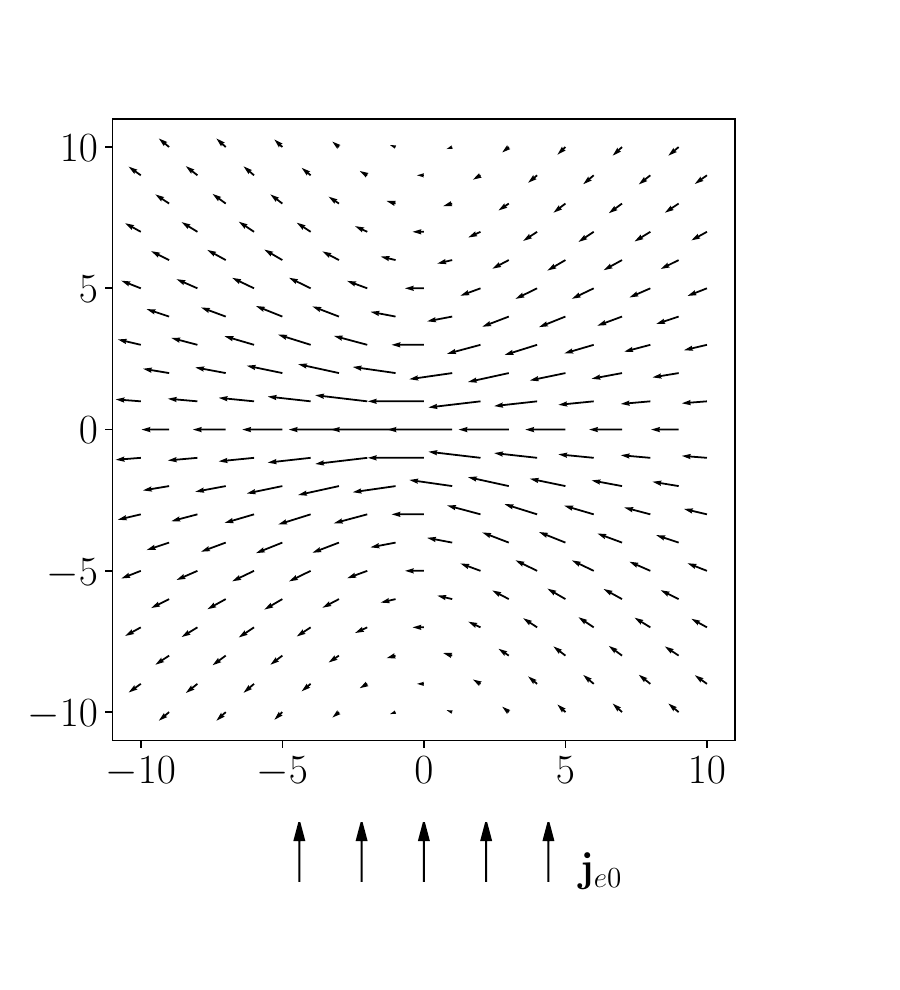}
        \hspace{1cm}
	\includegraphics[width=0.4\textwidth,trim= 0.8cm 1.5cm 2.7cm 1.5cm, clip = true]{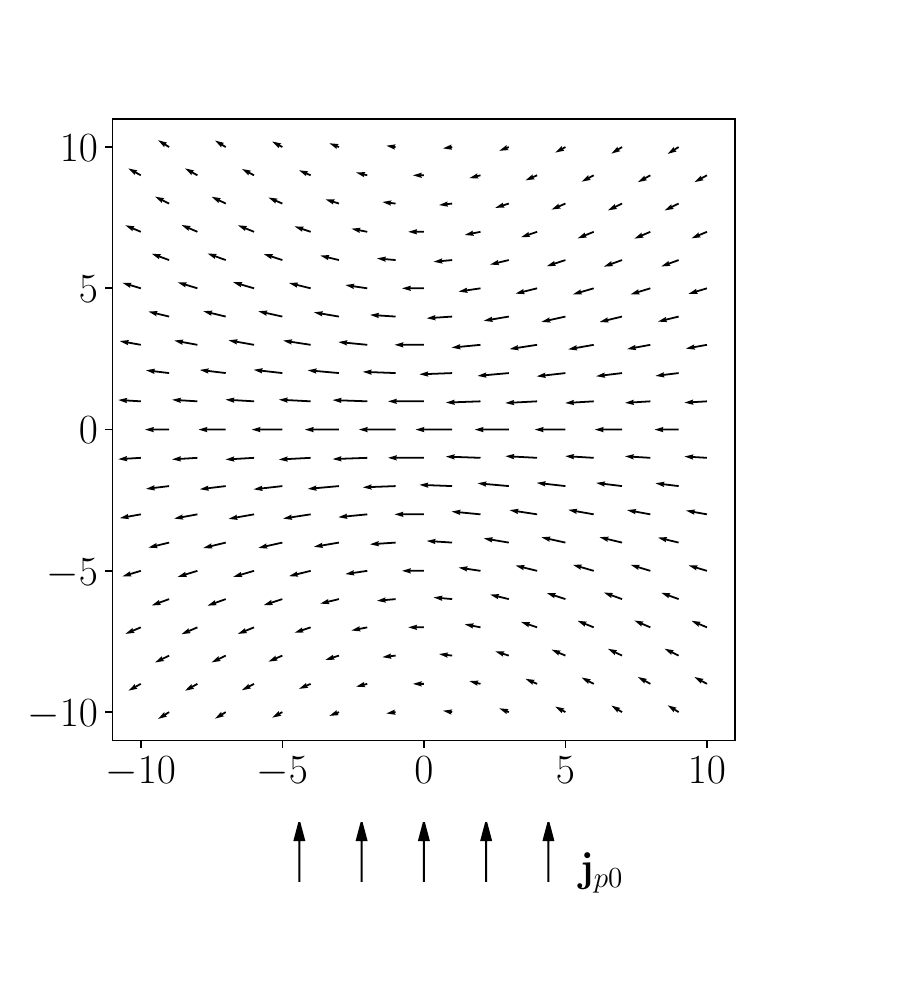}    
	\caption{
            \label{fig:dje_djp}
            The electron (left panel) and proton (right panel) particle current density corrections. 
            The coordinates, measured from the vortex axis, are normalized by the London penetration depth $\lambda$. The direction of the transport electron ($\vj_{e0}$)  and proton ($\vj_{p0}$) currents are indicated by vertical arrows (not to scale).
          }
    \end{figure}

    In the opposite limiting case of $\rho \gg \lambda$, we can utilize the representation of the function $\mathcal{P}'$ through the delta function, Eq.~\eqref{eq:ap:dP_approx}, obtaining
    \begin{equation}
        \int \limits_{-\infty}^{ \infty} dy_1 
        \frac{{\mathcal{P}}'\left(\sqrt{\rho^2 \sin^2\phi_p+y_1^2}\right)}{2 \pi\sqrt{\rho^2 \sin^2\phi_p+y_1^2}} \approx \delta (\rho \sin\phi_p) = \frac{\delta(\phi_p) + \delta(\phi_p+\pi)}{\rho}.
    \end{equation}
    According to this approximation, only the function $J_{\rho}(\rho)$ contributes at large distances,   
    leading to the current correction
    \begin{equation}
        \label{eq:dje_cv}    
        \delta \vj_e(\vrd) \approx - 2     \frac{3 m_p \varkappa}{16 p_{Fe}}\,  (\vj_{e0} \, \vec{e}_\phi) \, \frac{\vec{e}_\rho}{\rho}.
    \end{equation}    
    The behavior of $\delta \vj_e(\vrd)$ at intermediate distances is 
    illustrated in the left panel of Fig.~\ref{fig:dje_djp}.

    Finally, let us verify that the obtained solution satisfies the continuity equation, $\Div \vj_e = 0$.
    For the incident electron current $\vj_{e0}$, this is obvious. 
    Applying the operator $\Div$ to the correction $\delta \vj_e$, we get:
    \begin{equation}
        \Div \delta \vj_{e} = 
        -  ( \vj_{e0} \, \vec{e}_\phi) \left( \frac{d J_\rho}{d \rho} + \frac{J_\rho - J_\phi}{\rho} \right).
    \end{equation}
   In this expression, using the definitions \eqref{eq:J_rho} and \eqref{eq:J_phi}, we can transform the first term in the large parentheses as follows: 
   \begin{align}
        \frac{d J_\rho}{d \rho} = 
        & m_p \varkappa \frac{3}{16 p_{Fe}}  \int\limits_0^{2\pi} d\phi_p\cos^2\phi_p  \frac{\partial}{\partial \rho} \int \limits_{-\infty}^{ \infty} dy_1 \frac{{\mathcal{P}}'\left(\sqrt{\rho^2 \sin^2\phi_p+y_1^2}\right)}{2 \pi\sqrt{\rho^2 \sin^2\phi_p+y_1^2}}
        \\
        & =  m_p \varkappa \frac{3}{16 p_{Fe}}  \int\limits_0^{2\pi} d\phi_p \frac{\cos\phi_p \sin \phi_p}{\rho}   \frac{\partial}{\partial \phi_p} \int \limits_{-\infty}^{ \infty} dy_1 \frac{{\mathcal{P}}'\left(\sqrt{\rho^2 \sin^2\phi_p+y_1^2}\right)}{2 \pi\sqrt{\rho^2 \sin^2\phi_p+y_1^2}}
        \\
        & = - \frac{J_\rho - J_\phi}{\rho}.
    \end{align}
    Here, to derive the last equality, we performed the integration by parts. 
    Thus, we see that the correction $\delta \vj_{e}$ also satisfies the continuity equation.

\subsection{Protons}
\label{sec:protons}

    Let us now turn to the protons.
    To calculate the force on a vortex, we need to determine the proton particle current density.
   Specifically, we will need the expression for this current up to the terms linear in the incident flux, i.e., in the ratio
    \begin{equation}
        \label{eq:p_ineq}
        \frac{Q_{p0}}{p_{Fp}} \ll 1
    \end{equation}
    as well as linear in the ratio
    \begin{equation}
        \frac{Q_{pv}}{p_{Fp}} \ll 1.
    \end{equation}  
    The latter parameter cannot be considered  small in the strict sense since  $Q_{pv}$ diverges as $\rho \rightarrow 0$.
    As it was stated, we adopt a model where the vortex core is approximated by an infinitely thin line.
    However, in reality, the vortex core has a finite radius of the order of the coherence length $\xi$.
    Thus, in the region outside the core (where our model is valid), we can use Eqs.\ \eqref{eq:xi_def} and \eqref{eq:kappa_def} to estimate:
    \begin{equation}
        \label{eq:Qpv_est_n}
	\frac{{Q}_{pv}}{p_{Fp}} \lesssim \frac{m_p \varkappa}{2\pi \xi p_{Fp}} \sim\frac{\Delta^{(p)} m_p}{p_{Fp}^2} = \epsilon.
    \end{equation}
    The parameter $\epsilon$, formally introduced in Sec.~\ref{sec:scat_el}, naturally arises for protons.
    Strictly speaking, there is another dimensionless parameter in the problem:
    \begin{equation}
        \label{eq:another_small_p}
        \frac{Q_p p_{Fp}}{m_p \Delta^{(p)}_0},
    \end{equation}
    which cannot be considered small everywhere. 
    Finite values of this parameter lead to the existence of Bogoliubov excitations even at vanishing temperature \cite{Leinson2017}.
    This complicates the expression for the proton current density, reduces the proton gap compared to its unperturbed value
    $\Delta^{(p)}_0$,
    and eventually 
    results in
    the breakdown of superconductivity 
    \cite{Alexandrov2003,gk13,Leinson2017,AllardChamel2021b,AllardChamel2023}. 
    In our model, we include the region where these effects become substantial in the vortex core.

    Proceeding within the specified 
    level of accuracy, and taking into account representation \eqref{eq:Q_mod_ch}, the proton current density can be written as 
    \begin{equation}
	\label{eq:jp_rep}
	\vec{j}_p =   \vj_{p0} +   \vj_{pv} +  \delta \vj_{p},
    \end{equation}   
    where
    \begin{equation}
        \label{eq:jp0_def}
        \vj_{p0} =  \frac{n_{p0}}{m_p} \vQ_{p0}
    \end{equation}
    is the transport proton current (the proton current density measured far from the vortex),
    \begin{equation}
        \label{eq:jpv_def}
        \vj_{pv} =  \frac{n_{p0}}{m_p} \vQ_{pv}
    \end{equation}
    is the current generated by the vortex, and
    \begin{equation}
        \label{eq:djp_def}
        \delta \vj_{p} = \frac{n_{p0}}{m_p} \delta \vQ_{p}
    \end{equation}
    is the correction to the current to be discussed below. 
    In Eqs.\ \eqref{eq:jp0_def}--\eqref{eq:djp_def}, $n_{p0}$ is the unperturbed proton density, which is equal to
    the unperturbed electron density, $n_{e0}$ (the quasineutrality condition):
    \begin{align}
        n_{p0}=n_{e0}.
        \label{quasi}
    \end{align}
    One can notice that there are no terms arising due to the deviation of the proton number density from $n_{p0}$ in these expressions.
    The perturbation of a scalar  quantity, such as particle density, are proportional  to at least $Q_p^2$.
    Therefore, any corrections to the proton current arising from perturbations in the proton density would be at least quadratic in $Q_{p0}/p_{Fp}$ or in $\epsilon$.
    Consequently, including such terms would exceed the adopted level of accuracy.    
    Here, the correction $\delta \vQ_p$, being generated by the interaction of the incident flux with the vortex, is assumed to be at least linear in $Q_{p0}/p_{Fp}$ as well as in parameter $\epsilon$.

    Thus, all we need is to calculate the correction $\delta \vj_{p}$.
    To find the equation for  $\delta \vj_{p}$, we utilize Maxwell's equation \eqref{eq:Max_eq}.
    Substituting representations \eqref{eq:je_exp} and \eqref{eq:jp_rep} along with screening condition \eqref{eq:screen_cond_npe} and Eq.~\eqref{eq:Max_eq_v}, we obtain:
    \begin{equation}
	\label{eq:Max_eq_delta}
	\Rot \delta \vec{B} 
          = \frac{4 \pi e}{c}  \left( \delta \vj_p - \delta \vj_e \right),
   \end{equation}     
   where $\delta \vB $ is defined according to Eq.~\eqref{eq:B_def}.
   On the other hand, taking into account Eqs.~\eqref{eq:Q_def} and \eqref{eq:djp_def}, 
   we can relate $\Rot \delta \vj_{p}$ and $\delta \vB$:    
    \begin{equation}
        \label{eq:djp_rep}
        \Rot \delta \vj_{p} = 
        -  \frac{e}{c} \frac{n_{p0}}{m_p} \, \delta \vB.
    \end{equation}
    Substituting $\delta \vB$ from Eq.~\eqref{eq:djp_rep} into Eq.~\eqref{eq:Max_eq_delta}, we obtain:
    \begin{equation}
	\label{eq:Max_eq_delta2}
    	- \Rot \Rot \delta \vj_p
    = \lambda^{-2}  \left( \delta \vj_p - \delta \vj_e \right),   
    \end{equation}  
    where we used the definition of the London penetration depth \eqref{eq:lambda_def}.
    In view of the continuity equation, 
    \begin{equation}
        \label{eq:cont_eq_djp}
        \Div \delta \vj_p =0,
    \end{equation}
    the first term on the left-hand side can be represented as: $\Rot \Rot \delta \vj_p = - \Delta \delta \vj_p$.
    As a result, we arrive at the following equation,
     \begin{equation}
	\label{eq:djp_eq}
	\left(\Delta -\lambda^{-2} \right) \delta \vj_p =  - \lambda^{-2}  \delta \vj_e.
    \end{equation}
    which 
    can be solved with the Green's function method. 
    The corresponding Green's function is given by Eq.\ \eqref{eq:green_func}.
    Taking into account Eq.~\eqref{eq:dP_apr},
    the solution can be represented as
    \begin{equation}
        \label{eq:djp_reg_def}
        \delta \vj_{p}(\vrd)  =  \int d^2\vrd_1   \delta \vj_e(\vrd_1)  \frac{\mathcal{P}'(|\vrd_1 - \vrd|)}{2\pi |\vrd_1 - \vrd|}.
    \end{equation}
    Formally, the correction $\delta \vj_{p}$ is defined by Eq.~\eqref{eq:djp_reg_def} up to the general solution of the homogeneous Eq.\ \eqref{eq:djp_eq} with $\delta \vj_{e}=0$.
    As verified in  Appendix \ref{sec:backflow}, incorporating this solution does not influence the force on the vortex.
    Keeping this in mind, we will proceed with formula \eqref{eq:djp_reg_def} in what follows.

    At distances $\rho \gg \lambda$, according to relations \eqref{eq:ap:dP_approx}, the function $\mathcal{P}'(\rho)$ entering Eq.~\eqref{eq:djp_reg_def} can be approximated by the delta function. 
    Thus, at the large distances, we have simply:
    \begin{equation}
	\label{eq: dj_ld}
        \delta \vj_p  = \delta \vj_e,
    \end{equation}
    where the approximate expression for the correction $\delta \vj_e$ is given by Eq.~\eqref{eq:dje_ld}.
    In fact, this result can be derived directly from Eq.\ \eqref{eq:djp_eq} where, at the large distances,  
    operator $\Delta$ can be neglected compared to $\lambda^{-2}$.
    Hence, we see that, far from the vortex core, the total electric current vanishes, as expected for the superconducting mixture.

    Let us now consider the behavior of the proton current density correction in the vicinity of the vortex core.
    The current given by Eq.~\eqref{eq:djp_reg_def} varies on length-scales of the order of the London penetration depth. 
    Therefore, in the region $\rho \ll \lambda$, it is almost constant and can be estimated as
    \begin{equation}
        \delta \vj_{p}(\vrd) \approx \delta \vj_{p}(0) =  \int d^2\vrd_1   \delta \vj_e(\vrd_1)  \frac{\mathcal{P}'(\rho_1 )}{2\pi \rho_1}
        =   \frac{  \vec{e}_z \times \vj_{e0} }{2} \int d^2\vrd_1
        \left(J_\rho(\rho_1) + J_\phi(\rho_1) \right) \frac{\mathcal{P}'(\rho_1)}{2\pi\rho_1}. 
    \end{equation}
    Here, in the last equality, we made use of Eq.~\eqref{eq:dje_alt_repr}.
    Substituting the expressions \eqref{eq:J_rho} and \eqref{eq:J_phi} for the functions $J_\rho$ and $J_\phi$,
    and accounting for the screening condition \eqref{eq:screen_cond_npe}, we obtain the following result
    \begin{equation}
        \label{eq:djp_atline}
        \delta \vj_{p}(\vrd) \approx \delta \vj_{p}(0)  = m_p \varkappa \frac{3 \pi}{8 p_{Fe}} L_e^{-1} \vec{e}_z\times \vj_{p0},
    \end{equation}
    where we introduced the notation
    \begin{equation}
	\label{eq:Le_def}
	L_e^{-1} \equiv \frac{1}{2}\int\limits_{-\infty}^{\infty} dx \left( \int \limits_{-\infty}^{\infty} dy\frac{\mathcal{P}'(\sqrt{x^2+y^2})}{2\pi \sqrt{x^2+y^2}}   \right)^2.
    \end{equation}
    We observe that electron scattering off the vortex magnetic field generates a correction, $\delta \vj_{p}$, to the proton superfluid current density.
    This correction, as well as the electron correction to the current, $\delta \vj_{e}$,
    is regular in the vicinity of the vortex core and is directed perpendicular to the proton 
    transport current $\vj_{p0}$.
    For the function $\mathcal{P}'(\rho)$ given by Eq.\ \eqref{eq:dP_apr}, the quantity $L_e^{-1}$  equals
    \cite{Gusakov2019} 
    \begin{equation}
        \label{eq:Le_val}
        L_e^{-1} = 
        \frac{1}{8 \lambda}.
    \end{equation} 
    Substituting this expression into Eq.\ \eqref{eq:djp_atline} verifies that, in the vicinity of the vortex core, the proton current correction is precisely half of the electron current correction given by Eq.\ \eqref{eq:dje_cv}: $\delta \vj_{p}(\vrd)=\delta \vj_{e}(\vrd)/2$.
    The behavior of $\delta \vj_p(\vrd)$ at intermediate distances is 
    illustrated in the right panel of Fig.~\ref{fig:dje_djp}.

    The validity of the continuity equation for the correction $\delta \vj_p$ follows from that of the corresponding continuity equation for the correction $\delta \vj_e$. This becomes evident if one performs the variable change $\vrd_1 \rightarrow \vrd_1 + \vrd$ in Eq.~\eqref{eq:djp_reg_def}.

    The last thing we need to consider is the correction to the chemical potential, $\delta \breve{\mu}_p$.
    According to Eqs.\ \eqref{eq:mu_def} and \eqref{eq:mup_repr}, in our stationary problem, it can be expressed as 
    \begin{equation}
        \label{eq:mup_res}
        \delta \breve{\mu}_p = - e \phi.
    \end{equation}
    %

\section{Calculation of the force on the vortex}
\label{sec:force_calc}

    Now, we are well-equipped to calculate the force acting on the vortex. 
    Before proceeding, let us briefly discuss the approximations used in this calculation.
    Since the vortex velocity relative to the surrounding matter is assumed to be not too large, 
    we limit our analysis to the force calculated linearly with respect to the incident fluxes. 
    Another small parameter in the problem  is $\epsilon$ [see Eq.\ \eqref{eq:eps_p_def}].
    As we will see, linear accuracy in $\epsilon$ is insufficient, as it only captures the perpendicular component of the force.
    Therefore, we extend the calculation to include quadratic terms in $\epsilon$.

    We present the calculation of the force using two methods, introduced in Sec.~\ref{sec:force_bal}.
    The first one is based on the formula \eqref{eq:F_mv_def}.
    In the main part of the paper, we choose the cylinder's radius $R$ to be much larger than the London penetration depth. 
    In Appendix~\ref{sec:fin_dist_force}, we reproduce the calculations with the same method but for an arbitrary cylinder's radius, explicitly showing that the results are insensitive to the chosen value of $R$. 
    The second method employs the Magnus force formula~\eqref{eq:F_mv_Meth2}.
    We verify that both methods yield identical expressions for the force, consistent with the arguments laid out in Sec.\ \ref{sec:force_bal}.

\subsection{Method I } 
\label{sec:Method_I} 

    Let us begin by calculating the force using formula \eqref{eq:F_mv_def}.
    As argued in Sec.\ \ref{sec:force_bal}, the result should be independent of the cylinder's radius.
    It is most convenient to choose the radius $R$ to be much larger than $\lambda$ (surface $S_1$ in Fig.\ \ref{fig:volumes}).%
    %
    \footnote{Simultaneously, this radius should be much smaller than the intervortex distance $d_B$.}
    %

    The total stress tensor is the sum of three terms describing the contributions of the neutron-proton mixture, the electrons, and the electromagnetic field as given by Eqs.\ \eqref{eq:Pi_np_ap}, \eqref{eq:Pi_e_ap}, and \eqref{eq:Pi_EM_ap}, respectively. 
    We start with the electrons. 
    Substituting the distribution function \eqref{eq:Ne_ans} into the general expression \eqref{eq:Pi_e_ap} and extracting  the contributions linear in the incident fluxes, we get:
    \begin{equation}
	\label{eq:Pi_e_lin}
	\Pi_{e}^{ik} = \sum_{\vp\sigma}  p^i   \vg^k ( g_e - e \varphi )\frac{\partial \mN_{\vp,0 }^{ (e)}}{\partial \varepsilon_\vp^{(e)}}.
    \end{equation}	
    Here, we take into account that the electrostatic potential $\varphi$, being generated by the interaction of the incident particles with the vortex, is at least a linear function of $\vVre$ and $\vQ_{p0}$. 
    For the distances $\rho \gg \lambda$, one can make use of the approximate expression \eqref{eq:dpe_ld}.
    This allows us to write down the electron contribution to the force as 
    \begin{equation}
	\label{eq:Fmv_e}
	{F}_{\rm m \rightarrow v}^{(e)i} = -  \oint\limits_{\rho = R} \Pi_e^{ik} n_k dS 
	= - \frac{1}{2} \sum_{\vp\sigma} v_{\vp\perp} p_\perp^2 \left[ -( \vec{e}_z \times\vVre) \sigma_\perp^{(e)} + \vVre \sigma_{||}^{(e)} \right] \frac{\partial \mN_{\vp,0 }^{ (e)}}{\partial \varepsilon_\vp^{(e)}}
	+ e n_{e0} \int d^3 \vr \  \vec{E},
    \end{equation}
    where
    $v_{\vp\perp}  = p_\perp c^2 /\varepsilon_\vp^{(e)}$ and in the last term, we returned to the volume integration.

    Let us now consider the contribution from the neutron-proton mixture.
    The neutron condensate momentum coincides with that at infinity, $\vQ_n = \vQ_{n0}$, and cannot lead to any force on the vortex, since the neutron-related momentum flux through the cylinder's surface obviously vanishes.
    The case of proton superfluid flow is more complicated.
    Considering its contribution to the stress tensor \eqref{eq:Pi_np_ap}, we first note that at distances $\rho \gg \lambda$, the vortex azimuthal flow $\propto \vQ_{pv}$ is exponentially suppressed [see Eq.\ \eqref{eq:Q_pv_def} and the third property of the function $\mathcal{P}$  outlined after Eq.\ \eqref{eq:A}]. 
    Therefore, all the terms containing $\vQ_{pv}$ should be neglected.
    Among the remaining terms, the following combinations are present: $Q_{p0}^i \delta Q_{p}^k$, $\delta Q_{p}^i \delta Q_{p}^k$, and $Q_{p0}^i Q_{p0}^k$. 
    Since all these terms are quadratic in the incident flux, they should be omitted as they exceed the adopted level of accuracy.
    However, it can be verified that even beyond the accepted accuracy,
     each of these terms contributes negligibly to the force,
     either vanishing as $R \rightarrow \infty$ or canceling out after integration over the cylinder's surface.

    The final part of the tensor \eqref{eq:Pi_np_ap} to be considered is the pressure term.
    In view of Eq.\ \eqref{eq:dP}, we can write:
    \begin{equation}
        \label{eq:deltaP}
        \delta P_{np} = n_{n0} \delta \breve{\mu}_n + n_{p0} \delta \breve{\mu}_p
         = - e n_{p0} \, \varphi,
    \end{equation} 
    where the terms involving squared condensate momenta  are neglected for the same reasons as in the convective part of the stress tensor.
    To obtain the second equality, we substituted Eq.\ \eqref{eq:mup_res} for protons and noted that, according to Eq.\ \eqref{eq:neutron_sf_eq} for neutrons, we have $\delta \breve{\mu}_n = 0$. 
    As a result, the neutron-proton contribution to the total force reduces to 
    \begin{equation}
	\label{eq:Fmv_np}
	{F}_{\rm m \rightarrow v}^{(np)i}= -  \oint\limits_{\rho = R} \Pi_{np}^{ik} n_k dS 
        =
	- e n_{p0} \int d^3 \vr \  \vec{E}.
    \end{equation}

    Finally, let us consider the electromagnetic field contribution produced by the stress tensor \eqref{eq:Pi_EM_ap}.
    The electric field is linear in the incident fluxes; therefore, the electric part of the tensor is quadratically small.
    As for the magnetic part, in addition to the terms that are quadratically small in the incident fluxes, it contains terms proportional to $B_v^i B_v^k$ and $B_v^i \delta B^k$.
    However, these terms are exponentially suppressed at large distances and should also be neglected.
    Thus, there is no electromagnetic contribution within the accepted accuracy.

    Now, we can write down the total force acting on the vortex.
    First, we note that according to the quasineutrality condition \eqref{quasi}, the terms proportional to the electric field in Eqs.\ \eqref{eq:Fmv_e} and \eqref{eq:Fmv_np} cancel each other out.
    Let us expand the remaining part of the force into perpendicular and parallel components.
    Substituting Eq.\ \eqref{eq:sigma_e_perp} and integrating the perpendicular component, we obtain:
    \begin{equation}
	\label{eq:F_perp_meth_I}
	\vec{F}_{\rm m \rightarrow v \perp} = D'_e  \, \left( \vec{e}_z\times \vVre \right),
    \end{equation}
    where 
    \begin{equation}
        D'_e = - m_p \varkappa \, n_{e0}.
    \end{equation}

    Substituting Eq.\ \eqref{eq:sigma_e_par}  and integrating the parallel component of the force, we get: 
    \begin{equation}
	\label{eq:F_par_meth_I}
	\vec{F}_{\rm m \rightarrow v ||}
        = - D_e \, \vec{e}_z \times \left( \vec{e}_z \times \vVre \right),	 
    \end{equation}
    where 
    \begin{equation}
        \label{eq:De_expr}
        D_e = (m_p \varkappa)^2  \frac{3\pi}{8}   \frac{n_{e0}}{ p_{Fe}}  L_e^{-1}
    \end{equation}
    and 
    the quantity $L_e^{-1}$ is given by Eq.~\eqref{eq:Le_def}.
    In Sec.\ \ref{sec:scat_el}, we assumed that the electron velocity $\vVre$ is perpendicular to the $z$-axis.
    However, here we replace $\vVre$ with the equivalent vector $- \vec{e}_z \times \left( \vec{e}_z \times \vVre \right)$ to give the expression a more general form, accounting for the possible existence of a $z$-component in the velocity [cf.\ Eq.\ \eqref{eq:force_gen}].  
    Expression \eqref{eq:De_expr}, along with Eq.\ \eqref{eq:Le_val}, reproduces the result for the longitudinal force obtained in Ref.\ \cite{Gusakov2019} in the limit $\xi/\lambda \ll 1$.
    In that paper, the coefficient $D$ is given by the expression [see Eq.\ (60) there]:
    \begin{equation}
        D = m_p \varkappa \frac{3\pi}{8} \frac{\hbar n_{e0}}{p_{Fe}\xi} G(\lambda/\xi),
    \end{equation}
    where the function $G(\lambda/\xi)$, defined by Eq.\ (41) of Ref.\ \cite{Gusakov2019}, is approximately equal to $\pi\xi/8\lambda$ for $\xi/\lambda \ll 1$.
    Substituting this value and bearing in mind that $\pi\hbar = m_p \varkappa$, one can easily verify the equivalence of the results.
    We see that, at distances $R\gg \lambda$, the only source of the  force is the scattering of electrons off the vortex magnetic field. 

\vspace{0.2cm}

    \underline{\textit{Remark I}}. 
    In calculating the force, we used the fact that the electric field $\vec{E}$, generated by the interaction between the incident fluxes and the vortex, is at least a linear function of these fluxes. However, the exact form of the electric field was not required for our calculations and is therefore not presented in the paper.
    The expression for the corresponding electrostatic potential can be found in Ref.~\cite{Gusakov2019}, where, in particular,
     it is shown that  it decreases as $\rho^{-1}$ at large distances.

\vspace{0.2cm}

     \underline{\textit{Remark II}}.
    Matching the result of this section with the general expression given by Eq.~\eqref{eq:force_gen}, we conclude that $D_p = D'_p = D_n = D'_n = 0$.
    However, as noted in the Introduction, the definition of the coefficients $D_\alpha$ and $D'_\alpha$ is generally ambiguous.
    In the next section, an alternative representation of the force $\vec{F}_{\rm m \rightarrow v}$ is presented.

\subsection{Method II } 
\label{sec:Method_II}

    Let us now calculate the same force using Eq.~\eqref{eq:F_mv_Meth2}.
    First of all, we note that this expression contains the factor $m_p \varkappa = \pi^2 \xi p_{Fp} \epsilon$, i.e., it already includes the small parameter $\epsilon$.
    Therefore, to calculate the force with an accuracy of $\sim \epsilon^2$, we need 
    the proton current density up to terms of order $\sim \epsilon$.
    An appropriate expression was obtained in Sec.~\ref{sec:protons}.
    Substitution of the first term of Eq.\ \eqref{eq:jp_rep} gives the perpendicular component of the force: 
    \begin{equation}
        \label{eq:F_perp_meth_II}
	\vec{F}_{\rm m \rightarrow v, \perp} = 
	    - {m_p \varkappa}  \, \vec{e}_z \times \vj_{p0}.
    \end{equation} 
    This expression is the well-known Magnus force,
    which is perpendicular to the transport proton current $\vj_{p0}$.
    Taking into account the screening condition \eqref{eq:screen_cond_npe} and Eq.\ \eqref{eq:je0_def},   
    it is easy to see that Eq.\ \eqref{eq:F_perp_meth_II} coincides with Eq.\ \eqref{eq:F_perp_meth_I}.
    Thus, despite the different physical origins of the transverse force at large distances and near the vortex core, its magnitude remains the same, as expected (see Sec.\ \ref{sec:force_bal}).

    To derive the parallel component of the force, we must proceed further by substituting the remaining contributions, $\vj_{pv}$ and $\delta \vj_{p}$, from the proton current 
    \eqref{eq:jp_rep}
    into Eq.\ \eqref{eq:F_mv_Meth2}.
    Note that the vortex-generated current, $\vj_{pv} \propto \vQ_{pv}$, formally diverges at $\rho = 0$.
    To evaluate the two-dimensional integral in Eq.~\eqref{eq:F_mv_Meth2}, associated with $\vj_{pv}$, we first integrate over the azimuthal angle. 
    This leads to a vanishing force. 
    While this procedure is not entirely rigorous from a strict mathematical perspective since the integral's value  depends on the order of integration, the result is physically justified. 
    Indeed, the azimuthal vortex flow $\propto \vQ_{pv}$ lacks a preferred direction,  meaning it cannot define a specific direction for the force.

    The result above suggests that the longitudinal force arises from the correction to the proton current density, $\delta \vj_p$.
    Let us verify this.   
    We need to substitute $\delta \vj_p$ into Eq.~\eqref{eq:F_mv_Meth2} at the vortex line ($\vrd=0$).
    The required expression is given by Eq.~\eqref{eq:djp_atline}.
    In this way we obtain:
    \begin{equation}
                \label{eq:F_par_meth_II}
	\vec{F}_{\rm m \rightarrow v, ||} 
	=  - {m_p \varkappa} \  \int d^2 \rho \,  \frac{\delta(\rho)}{2\pi \rho} \vec{e}_z\times \delta \vj_{p}(\vrd)
    	 =  -  \left(m_p \varkappa \right)^2 \frac{3 \pi}{8 p_{Fe}} L_e^{-1} \vec{e}_z\times\left( \vec{e}_z\times \vj_{p0} \right).
    \end{equation}
    Again, taking into account the screening condition \eqref{eq:screen_cond_npe} with Eq.\ \eqref{eq:je0_def}, we arrive at Eq.\ \eqref{eq:F_par_meth_I}.  
    Thus, one can see that the longitudinal force indeed arises due to the proton current correction.
    The magnitude of this force, like that of its transverse counterpart, remains unchanged across different length scales, despite being driven by different physical mechanisms.
    Equations \eqref{eq:F_perp_meth_II} and \eqref{eq:F_par_meth_II} suggest an alternative (equivalent) representation of the force on a vortex in terms of the coefficients $D_\alpha$ and $D'_\alpha$ [see Eq.~\eqref{eq:force_gen}].
    In particular, we can define:%
    %
    \footnote{To match Eqs.~\eqref{eq:F_perp_meth_II} and \eqref{eq:F_par_meth_II} with the general expression~\eqref{eq:force_gen}, one should introduce the proton incident velocity $\vVrp = \vQ_{p0}/m_p$.  }
    %
    \begin{equation}
        D'_p = - m_p \varkappa \,  n_{p,0}, \ \ \ \ \ D_p = (m_p \varkappa)^2  \frac{3\pi}{8}   \frac{n_{p0}}{ p_{Fp}}  L_e^{-1},
    \end{equation}
    and set all other coefficients to zero: $D_e = D'_e = D_n = D'_n =0$.
    Expressing the force \eqref{eq:force_gen} in terms of the ``proton'' coefficients $D'_p$ and $D_p$ highlights the fact that protons are the sole transmitters of momentum directly to the vortex core (cf.\ {\it Remark II} in Sec.\ \ref{sec:Method_I}).

    As we saw, when calculating the force $\vec{F}_{\rm m \rightarrow v}$ using  Eq.~\eqref{eq:F_mv_Meth2}, one cannot immediately eliminate the delta function since $\vj_{pv}$ diverges as $\rho \rightarrow 0$.
    However, by applying a physically reasonable prescription, we obtain a finite value.
    The same procedure allows us to 
    present
    the force in the following familiar form:
    \begin{align}
        \vec{F}_{\rm m \rightarrow v}= - m_p \varkappa \, \vec{e}_z\times \vj_{ p}(0),
    \label{forceforce}
    \end{align}
    where, by definition, $\vj_{ p}(0)$  
    is given by
    \begin{equation}
        \label{eq:jp0_formagnus}
        \vj_{ p}(0) = \lim_{\rho\rightarrow 0} \int\limits_0^{2\pi}\frac{d \phi}{2\pi} \, \vj_p(\vrd).
    \end{equation}

    Summarizing the results of this section, we confirm through direct calculation that the total force derived using Method I (see also Ref.\ \cite{Gusakov2019}) is ultimately applied to the vortex core as the Magnus force, given by Eq.\ \eqref{forceforce}, with $\vj_{p}(0) = \vj_{p0} + \delta \vj_{p}(0)$.
    Specifically, the transverse force is governed by the transport proton current density, $\vj_{p0}$, while the correction $\delta \vj_{p}(0)$ is responsible for the longitudinal force.

\vspace{0.2cm}
    \underline{\textit{Remark}}.
    Since, according to Method II, momentum is transferred to the vortex core by protons (not electrons), it is natural to expect that calculating the same force using Method I, but with the cylinder's radius $R \ll \lambda$ (surface $S_2$ in Fig.~\ref{fig:volumes}), would lead to the same conclusion.
    In Appendix~\ref{sec:fin_dist_force}, we show how the momentum flux through the cylinder is gradually redistributed from electrons to protons as the radius $R$ decreases.
    When the cylinder's radius becomes much smaller than $\lambda$, Eq.\ \eqref{eq:F_mv_def} indeed reproduces the expression for the force given by Eq.\ \eqref{forceforce}, with the proton current defined according to Eq.\ \eqref{eq:jp0_formagnus} [see Eq.\ \eqref{eq:ap:Fmvp2_Magnus}].
    This result demonstrates the overall self-consistency of our calculation and further validates our method for handling the delta function in Eq.\ \eqref{eq:F_mv_Meth2}.

\section{Discussion and Summary}
\label{sec:summary}

    This paper presents a detailed  analysis of the forces acting on proton vortices in the $npe$-matter of NS, which consists of superfluid neutrons, superconducting protons, and electrons. The main results and conclusions are summarized as follows:

\begin{itemize}

    \item    
    An expression for the electron distribution function, accounting for their scattering by the vortex magnetic field, has been derived. The obtained expression is applicable at any distance from the vortex line.
    Using the derived electron distribution function, a correction to the proton transport current, arising as a response to electron scattering, has also been calculated.
    Electrons scattering off the vortex magnetic field transfer momentum to the protons that generate the field.
    Therefore, the correction to the proton current must be accounted for to ensure momentum conservation near the vortex during its interaction with the surrounding medium.

    \item 
    The force acting on the vortex can be determined using formula \eqref{eq:F_mv_def} as the momentum flux through the surface of a cylinder surrounding the vortex.
    It is explicitly demonstrated that the magnitude and direction of the resultant force are independent of the chosen radius of the cylinder (strictly speaking, the cylinder can be replaced by any closed surface of arbitrary shape).   

    If the cylinder's radius is much larger than the London penetration depth $\lambda$ (surface $S_1$ in Fig.\ \ref{fig:volumes}), the exclusive source of the force is the scattering of electrons by the vortex magnetic field \cite{Gusakov2019}.
    From this {\it distant} perspective, the vortex experiences a counterforce opposing the magnetic component of the Lorentz force acting on the incident electrons.

    In contrast, if the cylinder's radius is on the order of the coherence length (i.e., when considering the force applied directly to the vortex core; see surface $S_2$ in Fig.~\ref{fig:volumes}), the contribution of electrons is negligibly small.%
    \footnote{
        The contribution of electrons can be estimated as $\sim \Phi/\Phi_0 \sim \xi^2/\lambda^2$ \cite{Gusakov2019}, where $\Phi$ is the magnetic flux enclosed within a cylinder of radius $R \sim \xi$, while $\Phi_0 = c m_p \varkappa / e$ is the total magnetic flux associated with the proton vortex.
    }
    In this region, superconducting protons become the primary carriers of momentum toward the vortex core.
    At intermediate distances, the momentum flux is redistributed among electrons, protons, and the magnetic field.

    \item 
    The key result of the present study is that the only relevant force applied directly to the vortex core is the Magnus force exerted by the superconducting protons. 
    Typically, in the literature on vortex forces, the Magnus force is understood as the expression [see, e.g., Refs.\ \cite{nv66, donnelly05}, cf.\ Eq.\ \eqref{eq:F_perp_meth_II}]
    \begin{equation}
        {\vec F}_{\rm M, \, litt} = - m_p \varkappa \, \vec{e}_z \times \vj_{p0},
    \end{equation}
    where $\vj_{p0}$ is the transport superconducting current in the vortex rest frame, i.e., the current that exists far from the vortex (at distances much larger than $\lambda$).
    Accounting for the screening condition, Eq.\ \eqref{eq:screen_cond_npe}, 
    one verifies
    that this force matches the perpendicular component of the force induced by electron scattering off the 
    vortex magnetic field, Eq.\ \eqref{eq:F_perp_meth_I}.
    In the present work, it is demonstrated that, to ensure the transmission of the entire force -- including its longitudinal component \eqref{eq:F_par_meth_I} -- from length scales $\sim \lambda$ to the vortex core, one must allow for the deviation of the local proton current from the transport current $\vj_{p0}$ in the vicinity of the core.
    Technically, this can be achieved by replacing $\vj_{p0}$ with $\vj_{p}(0) = \vj_{p0} + \delta \vj_{p}(0)$ 
    in the familiar
    Magnus force formula, Eq.\ \eqref{forceforce}.
    It should be emphasized that, as expected, the Magnus force remains strictly perpendicular to the local current $\vj_{p}(0)$.
    However, the correction $\delta \vj_{p}(0)$ to the proton current, given by Eq.\ \eqref{eq:djp_atline}, being perpendicular to the transport proton current $\vj_{p0}$, produces a component of the force {\it parallel} to $\vj_{p0}$.
    To the best of our knowledge, the importance of the supercurrent correction in generating the longitudinal force component has not been previously demonstrated.

    \item 
    It is widely accepted (see, e.g., Refs.~\cite{SaulsSteinSerene1982,AlfordSedrakian2010})
    that, after averaging over a large number of vortices,%
    \footnote{This procedure relies on the assumption that the vortices are uncorrelated and that scattering events off individual vortices are independent.  This assumption appears to be physically justified when the average magnetic field  is much smaller than the first critical magnetic field, such that the vortices form a dilute ensemble.  }
    the longitudinal force $\vec{F}_{m \rightarrow v, ||}$ is responsible for  energy dissipation and subsequent heat release in the system.
    Both the mechanical energy of oscillations and, for example, magnetic energy can undergo dissipation \cite{gagl15}.
    Thus, we arrive at the nontrivial conclusion that the Magnus force can be responsible for dissipative phenomena in matter.
    This result, obtained for a relatively simple problem concerning the forces on a vortex in the $npe$-matter of NS may also be valid for other superconducting systems.

    \end{itemize}

    Finally, let us discuss the approximations and simplifications made in this  work.
    First, we considered only the simplest $npe$-composition of NS matter.
    However, incorporating other particle species, such as muons, is straightforward.
    The modifications required for Method I are discussed in Ref.\ \cite{Gusakov2019}.
    As for Method II, one simply needs to replace the screening condition \eqref{eq:screen_cond_npe} with $\vj_{p0} = \vj_{e0} + \vj_{\mu0}$ when substituting it into Eq.\ \eqref{eq:F_perp_meth_II}, and replace the correction $\delta \vj_{e}$ with $\delta \vj_{e} + \delta \vj_{\mu}$ in Eq.~\eqref{eq:djp_reg_def}, where the muon correction $\delta \vj_{\mu}$ is calculated in the same way as $\delta \vj_{e}$.

    Second, throughout the paper, we neglect the entrainment effect \cite{AndreevBashkin1976} between neutrons and protons.
    In Appendix \ref{sec:entr}, we show that incorporating this effect does not alter the results, apart from a redefinition of the London penetration depth $\lambda$.
    Next, we assume that the temperature is sufficiently low for proton and neutron thermal excitations to be negligible.
    Relaxing this assumption may affect the results in certain regimes.
    It is well known from studies of terrestrial superfluids that an incident normal flux generates additional forces, such as the Iordanskii force (the normal counterpart to the superfluid Magnus force) and the Kopnin-Kravtsov force, which arises from the scattering of incident thermal excitations off bounded quasiparticles \cite{SoninBook}.
    We expect the effect of quasiparticle scattering to be rather small due to the large mean free path compared to other length-scales in the problem (see, e.g., Ref.\ \cite{ShterninOfengeim2022}).%
    \footnote{The calculations in Ref.\ \cite{ShterninOfengeim2022} were performed for quasinucleons in normal matter. In the case of superfluid matter, the mean free paths become even larger due to the presence of a gap in the corresponding energy spectrum.}
    %
    As for the Iordanskii force, it is expected to be comparable in magnitude to the Magnus force at temperatures approaching the critical temperature of the phase transition.
    Thus, accounting for thermal excitations is a relevant subject for future work.

    The most significant simplification adopted in the present study is the assumption that the coherence length is infinitely small compared to the London penetration depth (the extreme type-II superconductivity limit).
    As demonstrated in Sec.\ \ref{sec:hierarchy}, this simplification is overly idealized for actual NS core conditions.
    Thus, for real proton vortices, we cannot conclude that the Magnus force is the only force acting on the vortex core, even in the low-temperature regime.
    In general, a finite portion of the momentum can be transferred to the vortex core by electrons scattered off the core's magnetic field.
    As a result, our conclusions cannot be directly applied to most of the NS bulk.
    Nevertheless,  we believe that the results obtained in this work are conceptually important since they contribute to a better understanding of the role of the Magnus force in the vortex dynamics.


\section{Acknowledgments}
The authors are grateful to D.~P.~Barsukov, E.~M.~Kantor, and P.~S.~Shternin  for useful comments and discussions.
The work was supported by the Ministry of Science and Higher Education of the Russian Federation under the state assignment FFUG-2024-0002 of the Ioffe Institute.

\appendix

\section{Proton superfluid equation in the presence of a vortex}
    \label{sec:p_sf_eq}

    In what follows, we use the notation $\partial_t \equiv\partial/\partial t$, $\partial_i \equiv \partial/\partial x_i$, and $[\hat{A},\hat{B}] \equiv \hat{A} \hat{B} - \hat{B} \hat{A}$, where $\hat{A}$ and $\hat{B}$ are two differential operators. 
    Since the vortex is present in the system ($\Rot \nabla \chi \neq 0$ along the vortex axis), we must be careful and assume that $[\partial_i,\partial_j] \neq 0$ and $[\partial_t,\partial_i] \neq 0$.
    Taking the gradient of Eq.\ \eqref{eq:mu_def} and the time derivative of Eq.\ \eqref{eq:Q_def}, and then combining the resulting equations, we obtain:
    \begin{equation}
        \frac{\partial\vQ_p}{\partial t } = - \nabla\breve{\mu}_p +e\vec{E}+ \frac{\hbar}{2}\,  [\partial_t, { \nabla}]\chi_p,
        \label{sfleqxxx}
    \end{equation}
    where the electric field $\vec{E}$ is defined according to  Eq.\ \eqref{El}.
    It can be seen that the resulting equation contains the commutator $[\partial_t, {\nabla}]\chi_p$.
    To evaluate this commutator, we must make an assumption about the functional dependence of the phase $\chi_p$ on the coordinates and time, i.e., we need to let the system ``know'' about the vortices. 
    For simplicity, let us assume that there is only a single straight vortex directed along the $z$-axis, and that the problem is $z$-independent.
    The vortex moves with velocity $\vv = \partial_t {\vrd}_L(t)$ in the laboratory frame, where $\vrd_L(t)$ is the cylindrical radius vector pointing to the vortex line position.
    Thus, the phase $\chi_p$ can be expressed as
    \begin{align}
        \chi_p=\chi_{p,  {\rm reg}}(\vrd, t)+\chi_{pv}(\vrd-\vrd_L(t)),
         \label{phase}
    \end{align}
    where $\chi_{p,  {\rm reg}}(\vrd, t)$ is the regular part of the phase in the sense that $[\partial_t, {\nabla}]\chi_{p,  {\rm reg}}(\vrd, t)=0$, while $\chi_{pv}(\vrd-\vrd_L(t))$ is the additional phase contribution arising due to the presence of the straight vortex, hence the dependence of $\chi_{pv}$ on $\vrd-\vrd_L(t)$.
    Using Eq.\ \eqref{phase}, the commutator $[\partial_t, {\nabla}]\chi_p$ can be represented as
    \begin{align}
        [\partial_t, \partial_i]\chi_p
        &=[\partial_t, \partial_i]\chi_{pv}(\vrd-\vrd_L(t))
        =
        \frac{\partial}{\partial \rho_{L}^j}\left(\partial_i \chi_{pv}\right) {V}_{L}^j
        -\partial_i \left( \frac{\partial \chi_{pv}}{\partial \rho_{L}^j} V_{L}^j\right)
        \nonumber
        \\
        & = -\partial_j \partial_i\chi_{pv} \, V_L^j + \partial_i(\partial_j\chi_{pv} \, V_L^j) =
        [\vv\times {\rm curl}{\nabla \chi_{pv}}]^i = [\vv\times {\rm curl}{\nabla \chi_{p}}]^i,
        \label{comm}
    \end{align}
    and Eq.\ \eqref{sfleqxxx} is transformed into its final form, 
    \begin{equation}
         \frac{\partial \vQ_p}{\partial t } = - {\nabla}\breve{\mu}_p +e\vec{E}+ \frac{\hbar}{2}\,  [\vv\times \Rot { \nabla \chi_{p}}],
     \label{sfleqxxx2}
    \end{equation}
    which coincides with Eq.\ \eqref{eq:sf_eq1}.
    Although this equation is derived for the simple case of a single straight vortex, the proposed derivation can be easily generalized to an arbitrary number of curved vortices.

\section{Momentum conservation}
    \label{sec:mom_cons}

    In this appendix, we derive the momentum conservation equation \eqref{eq:mom_eq_ap}.
    To this end, it is convenient to use the proton superfluid equation in the form of Eq.\ \eqref{eq:sf_eq2},  along with the continuity equation:
    \begin{equation}
        \label{eq:cont_eq_p_td}
        \frac{\partial n_p}{\partial t} + \Div  \vj_p  = 0,
    \end{equation}
    where the proton current density is given by Eq.\ \eqref{eq:jp_def}.
    A corresponding set of equations for neutrons takes the following form:   
    \begin{equation}
        \frac{\partial \vQ_n}{\partial t} + \frac{1}{m_n} ( \vQ_n \nabla ) \vQ_n + \nabla \left( \breve{\mu}_n - \frac{\vQ_n^2}{2 m_n}\right)
        =   0,
    \end{equation}
    \begin{equation}
        \label{eq:cont_eq_n_td}
        \frac{\partial n_n}{\partial t} + \Div   \vj_n  = 0,      
    \end{equation}  
    where the neutron current density is given by Eq.\ \eqref{eq:jn_def}.
    Combining all these equations, we obtain:
    \begin{align}
        \label{eq:ap:np_mom_eq}
        \frac{\partial \mathfrak{P}_{np}^i}{\partial t} 
        + \frac{\partial \Pi_{np}^{ik}}{\partial r^k}
         = 
         n_p\, f_{\rm Lp}^i
        - n_p\, f_{\rm M}^i,
    \end{align}
    where the forces on the right-hand side are given by Eqs.~\eqref{eq:Lorenz_def} and~\eqref{eq:Magnus_per_p}.
    In Eq.\ \eqref{eq:ap:np_mom_eq}, we introduced the neutron-proton momentum density 
    \begin{equation}
       \mathfrak{P}_{np}^i = n_p {Q}^i_p + n_n {Q}^i_n,
    \end{equation}
    the neutron-proton stress tensor 
    \begin{equation}
	\label{eq:Pi_np_ap}
	\Pi_{np}^{ik}  = \frac{n_p}{m_p} Q_p^i Q_p^k + \frac{n_n}{m_n} Q_n^i Q_n^k + \delta^{ik} P_{np},
    \end{equation}    
    and the neutron-proton pressure, whose differential is given by 
    \begin{equation}
        \label{eq:dP}
        d  P_{np}
        = n_p  \, d \left( \breve{\mu}_p - \frac{Q_p^2}{2 m_p} \right)
        + n_n \, d \left( \breve{\mu}_n - \frac{Q_n^2}{2 m_n} \right).
    \end{equation}

    To derive the electron momentum conservation equation, we multiply Eq.\ \eqref{eq:kin_eq_e} by the momentum $\vp$ and sum over the momenta and spin $\sigma$ states:
    \begin{equation}
        \label{eq:ap:e_mom_eq}
        \frac{\partial \mathfrak{P}_{e}^i}{\partial t}
        + \frac{\partial \Pi_{e}^{ik}}{\partial r_k}
        =
        n_e\, {f}_{\rm L e}^i.
    \end{equation}    
    Here,  we introduced the electron momentum density 
    \begin{equation}
        \mathfrak{P}_{e}^i = \sum_{\vp\sigma} p^i \mN^{(e)}_\vp,
    \end{equation}
    the electron stress tensor 
    \begin{equation}
	\label{eq:Pi_e_ap}
	\Pi_{e}^{ik} = \sum_{\vp\sigma}  p^i   \vg^k  \mN_{\vp}^{ (e)},
    \end{equation}
    and the Lorentz force per electron 
    \begin{equation}
        \vec{f}_{\rm L e} = - e  \left(\vec E + \frac{1}{n_e c} \vj_e \times \vB \right).
    \end{equation}

    Finally, by summing Eqs.\ \eqref{eq:ap:np_mom_eq} and \eqref{eq:ap:e_mom_eq} and expressing the total Lorentz force in the standard way  using Maxwell's equations (see, e.g., Ref.\ \cite{LL2}),  we arrive at  Eq.~\eqref{eq:mom_eq_ap}
    with the total momentum density given by
    \begin{equation}
        \label{eq:total_mom}
        \mathfrak{P}^i = \mathfrak{P}_{np}^i + \mathfrak{P}_{e}^i + \mathfrak{P}_{\rm EM}^i,
    \end{equation}
    and the total stress tensor 
    \begin{equation}
        \label{eq:total_stress}
         \Pi^{ik}  = \Pi_{np}^{ik}  +  \Pi_{e}^{ik} +  \Pi_{EM}^{ik}.
    \end{equation}
    In these expressions, $\mathfrak{P}_{\rm EM}$ is the electromagnetic momentum density,
    \begin{equation}
        \mathfrak{P}_{\rm EM}^i = \frac{(\vec{E}\times\vec{B})^i}{4\pi c},
    \end{equation}
    and $\Pi_{\rm EM}^{ik}$ is the electromagnetic stress tensor,
    \begin{equation}
	\label{eq:Pi_EM_ap}
	\Pi_{\rm EM}^{ik} = - \frac{1}{4\pi} \left[ E^iE^k + B^i B^k - \frac{\delta^{ik}}{2}\left( E^2 + B^2\right) \right].
    \end{equation}    
    %

\section{Solution to Eqs.\ \eqref{eq:dpe1_eq} and  \eqref{eq:dpe2_eq}} 
\label{sec:sol_to_e}

    Let us define a frame of reference in which the electron velocity vector $\vvg$ (as well as the momentum vector $\vp$) lies in the $yz$-plane.
    In this case, assuming that the function $g_{e}$ does not depend on the $z$-coordinate, we can directly obtain 
    from Eq.\ \eqref{eq:dpe1_eq}:
    \begin{equation}
		\label{eq:dvpe1_sol_ap}
		g_{e1} = - \frac{  m_p \varkappa }{2 \pi }  \vec{e}_{\vp_\perp}  \left( \vec{e}_z \times \vVre \right) \int \limits_{-\infty}^{ y}  dy_1  \frac{\mathcal{P}'\left(\sqrt{x^2+y_1^2}\right)}{\sqrt{x^2+y_1^2}} 
		+ \mathcal{C}(x),
	\end{equation}
    where $\vec{e}_{\vp_\perp} = \vp_\perp/\vp_\perp$ and $ \mathcal{C}(x)$ is some function of $x$-coordinate. 
    From the requirement that the function $g_{e}$ vanishes at infinity, it follows that $\mathcal{C}(x) = 0$.
    \begin{figure}
	\includegraphics[width=0.4\textwidth,trim= 3.5cm 3.5cm 3.0cm 1.5cm, clip = true]{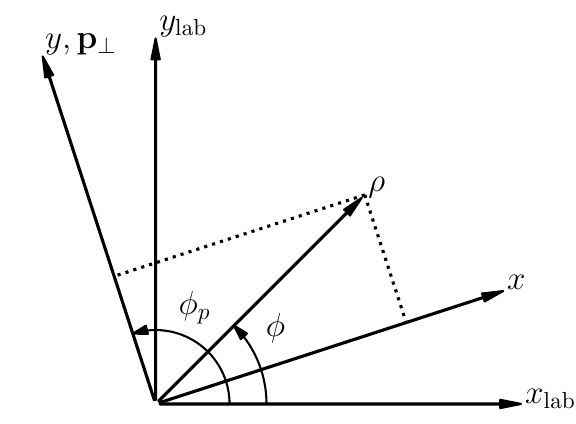}
        \caption{\label{fig:coords_2d}
           The coordinates and angles used in this paper.
           ${x}_{\rm lab}$ and ${y}_{\rm lab}$ denote the coordinates in a space-fixed frame, while $x$ and $y$ represent the coordinates in the frame attached to the vector $\vp_\perp$ (with the axis $y$ directed along $\vp_\perp$). The angles $\phi$ and $\phi_{\vp}$ are defined as shown in the figure.
            } 
    \end{figure}
    The next step is to substitute Eq.\ \eqref{eq:dvpe1_sol_ap} into Eq.\ \eqref{eq:dpe2_eq}.
    Before doing so, recall that the spatial coordinates $x$ and $y$ are defined relative to the vector $\vp$ (see Fig.\ \ref{fig:coords_2d}). 
    Therefore, they should be considered as functions of the angle 
    $\phi_p$,
    i.e.,
    \begin{equation}
	   \label{eq:xy_def_ap}
		x = \rho \sin(\phi_p - \phi), \ \ \ \  y = \rho \cos(\phi_p - \phi)
	\end{equation}
    and, consequently, 
    \begin{equation}
		\label{eq:mn1_el_ap}
		g_{e1} = - \frac{  m_p \varkappa }{2 \pi }  \vec{e}_{\vp_\perp}  \left( \vec{e}_z \times \vVre \right) \int \limits_{-\infty}^{ \rho \cos(\phi_p - \phi)}  dy_1  \frac{\mathcal{P}'\left(\sqrt{\rho^2 \sin(\phi_p - \phi)^2+y_1^2}\right)}{\sqrt{\rho^2 \sin(\phi_p - \phi)^2+y_1^2}}.
    \end{equation}
    Taking these identities into account, we can calculate the momentum derivative:
    \begin{align}
	\nonumber
	   \left[ \left(\vp \times \vec{e}_z\right)  \frac{\partial }{\partial \vp} \right]  g_{e1} 
		 &= \frac{  m_p \varkappa }{2 \pi }   \left\{ \vec{e}_{\vp_\perp} \vVre    \int \limits_{-\infty}^{ y} dy_1 \frac{\mathcal{P}'\left(\rho_1\right)}{\rho_1} 
	 	-  \vec{e}_{\vp_\perp}  \left( \vec{e}_z  \times \vVre\right) 
	 		\left[x \frac{\mathcal{P}'\left(\rho\right)}{\rho}
			- x y \int \limits_{-\infty}^{ y}  \frac{dy_1}{\rho_1}  \frac{d}{d\rho_1}\left(\frac{\mathcal{P}'\left(\rho_1\right)}{\rho_1}	\right)	
			\right]
	 	\right\}
            \\
            \color{red}
            &= \frac{  m_p \varkappa }{2 \pi }   \left\{ \vec{e}_{\vp_\perp} \vVre    \int \limits_{-\infty}^{ y} dy_1 \frac{\mathcal{P}'\left(\rho_1\right)}{\rho_1} 
	 	+  \vec{e}_{\vp_\perp}  \left( \vec{e}_z  \times \vVre\right) 
			 x\, y \int \limits_{-\infty}^{ y}  \frac{dy_1}{y_1^2} \frac{\mathcal{P}'\left(\rho_1\right)}{\rho_1}	
	 	\right\},
		\label{eq:ap:ddp_dp}
    \end{align}
    where we defined $\rho_1 = \sqrt{x^2+y_1^2}$.
    Substituting Eq.\ \eqref{eq:ap:ddp_dp} into \eqref{eq:dpe2_eq} and performing the integration, we obtain:
    \begin{align}
		\label{eq:mn2_el_ap}
		   g_{e2} &=  \frac{(m_p \varkappa)^2 }{4 \pi^2}  \frac{1}{p_\perp}
		    \left\{
		  	\frac{1}{2} \vec{e}_{\vp_\perp} \vVre   \left(  \int \limits_{-\infty}^{ y} dy_1
			   \frac{\mathcal{P}'\left(\rho_1\right)}{\rho_1} 
			\right)^2 
        +  \vec{e}_{\vp_\perp} \left( \vec{e}_z \times \vVre \right) x 
		    \int \limits_{-\infty}^{ y}dy_1  y_1 \frac{\mathcal{P}'\left(\rho_1\right) }{\rho_1}  
	 \int \limits_{-\infty}^{y_1} \frac{dy_2}{y_2^2} \frac{\mathcal{P}'\left(\rho_2\right) }{\rho_2}  
		  \right\},
    \end{align}	
    where $\rho_2 = \sqrt{x^2+y_2^2}$ and we used the fact that for a given function $S(y)$ the following identity is fulfilled:
    \begin{equation}
		\label{eq:ap:int_iden}
		\int_{-\infty}^{y} dy_1\, S(y_1) \int_{-\infty}^{y_1} dy_2 \, S(y_2) = \frac{1}{2} \left(\int_{-\infty}^{y} dy_1\,  S(y_1)\right)^2.
    \end{equation}

    The obtained solution is valid for any distance within the range $\xi \ll \rho \ll d_B$.
    However, considering that the function $\mathcal{P}(\rho)$ has a characteristic length-scale $\sim \lambda$, we can derive an approximate form of Eqs.\ \eqref{eq:mn1_el_ap} and \eqref{eq:mn2_el_ap} for distances $\rho \gg \lambda$.
    Let us first consider the function $g_{e1}$ given by Eq.\ \eqref{eq:mn1_el_ap}.
    At length scales much larger than $\lambda$, we can approximate: 
    \begin{equation}
        \frac{\mathcal{P}'\left(\sqrt{x^2+y^2}\right)}{2\pi \sqrt{x^2+y^2}} \approx \alpha_1 \, \delta(x)\delta(y),
    \end{equation}
    where the parameter $\alpha_1$ is determined 
    by ensuring that the integration of both sides over the entire area yields the same result.
    Using the third property of the function $\mathcal{P}$ (see Sec.\ \ref{sec:basic_eqs}), we can readily verify that $\alpha_1 = 1$.
    Hence, we have:
    \begin{equation}
        \label{eq:ap:dP_approx}
        \frac{\mathcal{P}'\left(\sqrt{x^2+y^2}\right)}{2\pi \sqrt{x^2+y^2}} \approx \delta(x)\delta(y) 
        =\frac{\delta(\rho)}{2\pi\rho},
    \end{equation}    
    where relation \eqref{eq:deltas_rel} was used.
    Furthermore, if the coordinates are defined according to Eqs.\ \eqref{eq:xy_def_ap},  the following identities can be established:
    \begin{equation}
        \int\limits_{-\infty}^{\infty} dy \, \delta(x)\delta(y) = \delta(x)\theta(y)  = \frac{\delta(\phi - \phi_p)}{\rho},
    \end{equation}
    where $\theta(y)$ is the Heaviside step function.
    As a result, the function $g_{e1}$ can be approximated as
    \begin{equation}
	g_{e1} \approx - {  m_p \varkappa }   \,  \vec{e}_{\vp_\perp} \left( \vec{e}_z \times \vVre \right)   \frac{\delta(\phi - \phi_p)}{\rho}.
    \end{equation}

    Let us now turn to the function $g_{e2}$ given by Eq.\ \eqref{eq:mn2_el_ap}.
    First, we  notice that the second term in Eq.\ \eqref{eq:mn2_el_ap}, which is
    directed along $\vp_\perp\times\vec{e}_z$, contains the factor $x$ and, consequently, vanishes when integrated  over $x$. 
    Therefore, this term can be neglected.
    Next, instead of Eq.\ \eqref{eq:ap:dP_approx} we have:
    \begin{equation}
        \frac{\mathcal{P}'\left(\sqrt{x^2+y^2}\right)}{2\pi \sqrt{x^2+y^2}}   \int \limits_{-\infty}^{ y} dy_1
			   \frac{\mathcal{P}'\left( \sqrt{x^2+y_1^2} \right)}{2\pi \sqrt{x^2+y_1^2}} \approx \alpha_2 \, \delta(x) \delta(y),
    \end{equation}
    where we used identity \eqref{eq:ap:int_iden}.
    Accounting for proper normalization with  the coefficient $\alpha_2$, we arrive at  
    \begin{equation}
        g_{e2} \approx 
	  \vec{e}_{\vp_\perp} \vVre  \, \frac{(m_p \varkappa)^2 }{p_\perp}  \frac{1}{8\pi^2}\int\limits_{-\infty}^{\infty} dx \left( \int \limits_{-\infty}^{\infty} dy\frac{\mathcal{P}'(\sqrt{x^2+y^2})}{\sqrt{x^2+y^2}}   \right)^2  \frac{\delta(\phi-\phi_p)}{\rho}.
    \end{equation}

\section{Vortex magnetic field} 
\label{sec:mag_field}

    Let us determine the magnetic field generated by a vortex \cite{deGennes_book}.
    The magnetic field should satisfy the Maxwell's equation 
    \begin{equation}
	\label{eq:Max_eq}
	\Rot \vec{B} = \frac{4 \pi e}{c} (\vj_p - \vj_e).
    \end{equation}
    In this appendix, we assume that there are no flows at large distances and, consequently, no corrections to the currents generated by scattering off the vortex. 
    Thus, for electrons, we set
    $\vj_e  = 0$, and Maxwell's equation becomes:
    \begin{equation}
        \label{eq:Max_eq_v}
        \Rot \vec{B}_v =   \frac{4 \pi e}{c}  \vj_{pv},
    \end{equation}
    where the proton current density is given by Eq.\ \eqref{eq:jpv_def}.
    Taking $\Rot$ of Eq.\ \eqref{eq:Max_eq_v}, substituting Eq.\ \eqref{eq:rotQ_ch}, and using the fact that $\Div \vec{B} = 0$, we obtain: 
    \begin{equation}
	\label{eq:Bv_eq}
	\left(\Delta - \lambda^{-2}	\right) \vec{B}_v = - m_p \varkappa  \frac{4 \pi e}{c} \frac{\delta(\rho)}{2\pi \rho} \vec{e}_z,
    \end{equation}
    where the London penetration depth is given by Eq.\ \eqref{eq:lambda_def}.
    Taking into account identity \eqref{eq:deltas_rel}, we note that Eq.\ \eqref{eq:Bv_eq} has the form of a two-dimensional Green's function equation:
    \begin{equation}
	\left(\Delta_{\pmb \rho} - \lambda^{-2}	\right)G(|\vrd - \vrd'|)   = \delta^{(2)}(\vrd - \vrd').
    \end{equation}
    The solution to this equation is well known:
    \begin{equation}
	\label{eq:green_func}
	G(|\vrd - \vrd'|)  = - \frac{1}{2\pi} K_0 \left(\frac{|\vrd - \vrd'|}{\lambda}\right).
    \end{equation}
    Using it, we arrive at Eq.\ \eqref{eq:B_vortex}, where the function $\mathcal{P}'(\rho)$ is given by Eq.\ \eqref{eq:dP_apr}.

\section{Forces at arbitrary distances}
    \label{sec:fin_dist_force}

    In this appendix, we calculate the electron, proton, and electromagnetic field contributions to the total force at an arbitrary distance $R$ from the vortex line.
    Having obtained these contributions, we show that the resultant force is independent of distance and corresponds to the Magnus force applied to the vortex core.
    Strictly speaking, this conclusion follows directly from the momentum conservation equation \eqref{eq:mom_eq_ap}.
    However, the explicit expressions provide valuable insights into how the various forces evolve with distance.

    Let us start with the proton contribution.
    Substituting the neutron-proton stress tensor \eqref{eq:Pi_np_ap} and the representation \eqref{eq:Q_mod_ch} into formula \eqref{eq:F_mv_def},  while keeping terms accurate to linear order in the incident fluxes and quadratic order in the small parameter $\epsilon$, we obtain:
    \begin{equation}
        \label{eq:ap:Fmvp1}
        \vec{F}_{\rm m \rightarrow v }^{(p)}     
        = - \frac{n_{p0}}{m_p} R \int \limits_0^{2\pi} d \phi \, \vQ_{pv} [(\vQ_{p0} + \delta \vQ_p) \vec{e}_\rho   ]    
        - R \int \limits_0^{2\pi} d\phi \, \delta P_{np} \,  \vec{e}_\rho  .
    \end{equation}
    To extract the terms linear in the incident flux from the pressure correction $\delta P_{np}$, we use identity \eqref{eq:dP}, where we further substitute Eq.\ \eqref{eq:mup_res} and set $\delta \breve{\mu}_n = 0$.
    Then, after some algebraic transformations, we obtain:
    \begin{equation}
        \label{eq:ap:Fmvp2}
        \vec{F}_{\rm m \rightarrow v }^{(p)}     
        = - m_p \varkappa  \int \limits_0^{2\pi} \frac{d\phi}{2\pi} \vec{e}_z \times  \left[\vj_{p0} + \delta \vj_p(R,\phi) \right] [1-\mathcal{P}(R)]
        - e n_{p0} \int\limits_{S_R} d^2\rho\, \vec E .        
    \end{equation}
    In this equation, we also substituted Eqs.\ \eqref{eq:Q_pv_def}, \eqref{eq:jp0_def}, and \eqref{eq:djp_def}. 
    The integration in the last term is carried out over a circular area $S_R$ with radius $R$.

    Considering the limiting case of $R \to 0$, we note that the function $\mathcal{P}(R)$ tends to zero (see Sec.\ \ref{sec:basic_eqs}).
    The last term in Eq.~\eqref{eq:ap:Fmvp2}  also tends to zero.
    Thus, Eq.~\eqref{eq:ap:Fmvp2} reduces to
    \begin{equation}
        \label{eq:ap:Fmvp2_Magnus}
        \vec{F}_{\rm m \rightarrow v }^{(p)}     
        = - m_p \varkappa  \int \limits_0^{2\pi} \frac{d\phi}{2\pi} \vec{e}_z \times  \left[\vj_{p0} + \delta \vj_p(R,\phi) \right] 
    \end{equation}  
    It is easy to see that this equation reproduces the expression for the force given in Eq.\ \eqref{forceforce}, which was obtained using Method II.

    In the opposite limiting case of $R \to \infty$ (i.e., when $\mathcal{P}(R) \to 1$), the first term in Eq.~\eqref{eq:ap:Fmvp2} tends to zero.
    Then, the only contribution to the total force at large distances from the neutron-proton mixture originates from the electrostatic interaction with the induced vortex electric field (the second term in Eq.\ \eqref{eq:ap:Fmvp2}, cf.\ Sec.\ref{sec:Method_I}).
    However, as we will see, this force cancels out with the corresponding term in the electron contribution.

    The first term in Eq.\ \eqref{eq:ap:Fmvp2}, by its derivation, represents the Magnus force acting on a cylinder of radius $R$ (see, e.g., Ref.\ \cite{SoninBook}).
    Indeed, this force is caused by the momentum flux generated by the interaction of the proton circulation flow with the proton incident flux.
    However, in contrast to uncharged fluids, where the circular flow drops with distance as $1/\rho$ and, consequently, the strength of the force is independent of the cylinder's radius, here, the circular flow $\propto \vQ_{pv}$ decreases exponentially (see Sec.~\ref{sec:basic_eqs}).
    This leads to the exponential suppression of this Magnus-like force with distance.

    One can formally interpret the same force in an alternative way. Using Eq.\ \eqref{eq:ap:np_mom_eq} (where we set the time derivative to zero and assume the vortex is at rest) and applying Gauss’s theorem, we can represent $\vec{F}_{\rm m \rightarrow v }^{(p)}$ as
    \begin{equation}
        \vec{F}_{\rm m \rightarrow v }^{(p)} =
            \vec{F}_{\rm{M}} + \vec{F}_{\rm{LM}p} + \vec{F}_{\rm{LE}p},
    \end{equation}
    where 
    \begin{equation}
        \label{eq:FMp}
        \vec{F}_{\rm{M}} =- m_p \varkappa  \int\limits_{S_R} d^2  \rho \,   \delta^{(2)}(\pmb \rho ) \ \vec{e}_z \times \left( \vj_{p0} + \delta \vj_p \right),
    \end{equation}
    \begin{equation}
        \vec{F}_{\rm{LM}p} =   m_p \varkappa  \int\limits_{S_R} d^2 \rho \, \left[ \vec{e}_z \times \left( \vj_{p0} + \delta \vj_p \right) \,  \frac{\mathcal{P}'(\rho)}{2 \pi \rho} 
        + \frac{e}{c}  \delta \vB  \times \vec{e}_\phi \,  \frac{ 1 - \mathcal{P}(\rho)}{2 \pi \rho} \right],
    \end{equation}
    and 
     \begin{equation}
     \label{prot11}
        \vec{F}_{\rm{LE}p} =    - e n_{p0} \int\limits_{S_R} d^2\rho\, \vec E.
    \end{equation}
    To obtain these expressions, we substituted representations \eqref{eq:Q_def} and \eqref{eq:B_def} for the total condensate momentum and magnetic field.
    Then, maintaining the level of accuracy adopted in this paper, we employed Eqs.\ \eqref{eq:B_vortex}, \eqref{eq:Q_pv_def}, \eqref{eq:jp0_def}, and \eqref{eq:djp_def}.
    Here, $\vec{F}_{\rm{M}}$ can be recognized as the standard distance-independent Magnus force, while $\vec{F}_{\rm{LM}p}$ and $\vec{F}_{\rm{LE}p}$ represent the magnetic and electric components of the counterforce to the Lorentz force.
    It is important to note, however, that $\vec{F}_{\rm{M}}$ is applied to the infinitely thin vortex core, whereas the forces $\vec{F}_{\rm{LM}p}$ and $\vec{F}_{\rm{LE}p}$ are applied to the bulk of the cylinder of radius $R$.

    Let us now turn to the electron contribution.
    To calculate the force, we again substitute the corresponding stress tensor into Eq.\ \eqref{eq:F_mv_def} and make use of Eq.\ \eqref{eq:ap:e_mom_eq} (with the time derivative set to zero), which, together with Gauss's theorem, yields the following expression:
    \begin{equation}
        \label{eq:ap:Fmve}
        \vec{F}_{\rm m \rightarrow v }^{(e)} 
        =   \vec{F}_{\rm{LM}e} + \vec{F}_{\rm{LE}e},
    \end{equation} 
    where 
    \begin{equation}
        \label{eq:FLMe}    
        \vec{F}_{\rm{LM}e} =  - m_p \varkappa  \int\limits_{S_R} d^2\rho \, \vec{e}_z \times  \left(\vj_{e0} + \delta \vj_e \right)  \frac{\mathcal{P}'(\rho)}{2 \pi \rho},
    \end{equation}
    and
    \begin{equation}
        \label{eq:FLEe}
        \vec{F}_{\rm{LE}e} = e n_{e0} \int\limits_{S_R} d^2\rho\, \vec E.
    \end{equation}
    Here, we again retained terms with the same level of accuracy as for the protons and substituted Eq.\ \eqref{eq:B_vortex}.

    As $R \rightarrow 0$, both the terms in Eq.~\eqref{eq:ap:Fmve} tend to zero.
    Therefore, there is no electron contribution to the total force acting on a cylinder with a radius much smaller than the London penetration depth. 
    In the opposite limiting case of $R \rightarrow \infty$, the expression for the force  reduces to the result obtained in Sec.\ \ref{sec:Method_I}.
    The equivalence between \eqref{eq:FLEe} and the last term of Eq.\ \eqref{eq:Fmv_e} at $R\rightarrow \infty$ is immediately evident. 
    Substitution of Eq.\ \eqref{eq:je0_def} into Eq.\ \eqref{eq:FLMe} gives us the transverse force \eqref{eq:F_perp_meth_I}. 
    To verify this, one just needs to use the first and the third properties of the function $\mathcal{P}(\rho)$ (see Sec.\ \ref{sec:basic_eqs}).
    Finally, the term containing electron current density correction $ \delta \vj_e$, is equivalent to Eq.\ \eqref{eq:F_par_meth_II}, which gives the longitudinal force (see Sec.\ \ref{sec:Method_II}).
    In summary, as $R\rightarrow \infty$, the force on a vortex is determined by the electron contribution (\ref{eq:FLMe}), while the only non-vanishing proton contribution  (\ref{prot11}) cancels out a similar electron term (\ref{eq:FLEe}).

    It remains to consider the contribution from the electromagnetic field.
    As argued in Sec.\ \ref{sec:Method_I}, there is no contribution from the electric field at the adopted level of accuracy.
    Substituting the magnetic part of the stress tensor \eqref{eq:Pi_EM_ap} into formula \eqref{eq:F_mv_def} and neglecting terms quadratic in the incident fluxes, we obtain:
    \begin{equation}
        \label{eq:FEM1}
        \vec{F}_{\rm m \rightarrow v }^{(EM)} = -\frac{R}{4 \pi}  \int \limits_0^{2\pi} d \phi \, (\vB_v \delta \vB) \vec{e}_\rho 
        = - m_p\varkappa \frac{c}{e}  \frac{R}{4 \pi}  \frac{\mathcal{P}'(R)}{2 \pi R} \int \limits_0^{2\pi} d \phi \,  \delta B  \,\vec{e}_\rho .
    \end{equation}
    Here, we took into account that both $\vB_v$ and $\delta \vB$ are directed along the $z$-axis and substituted Eq.\ \eqref{eq:B_vortex} in the second equality.
    This force tends to zero as $R \to 0$, since $\delta \vB$  is a regular function in the vicinity of the vortex core for the considered solution, Eq.~\eqref{eq:djp_reg_def}. 
    The vanishing of the force can be confirmed either by considering the second property of the function $\mathcal{P}$ (see Sec.\ \ref{sec:basic_eqs}) or by directly verifying that $\mathcal{P}'(R) \to 0$ using the explicit expression \eqref{eq:dP_apr}.
    According to the third property, $\mathcal{P}'(R)$ also tends to zero as $R \to \infty$. Thus, the electromagnetic contribution vanishes both in the vicinity of the vortex core and at large distances.
    Applying again the Gauss's theorem, using Eqs.\ \eqref{eq:Max_eq_v} and \eqref{eq:Max_eq_delta}, and substituting Eqs.\ \eqref{eq:B_vortex}, \eqref{eq:Q_pv_def}, \eqref{eq:jpv_def}, and \eqref{eq:djp_def}, we obtain:%
    %
    \footnote{Here, one should also make use of the vector analysis identity: $\nabla (\vec{a} \vec{b}) = (\vec{a} \nabla) \vec{b} + (\vec{b} \nabla) \vec{a} + \vec{a}\times \Rot\vec{b} + \vec{b}\times \Rot\vec{a} $, where $\vec{a}$ and $\vec{b}$ are two arbitrary vector fields (in our case, $\vB_v$ and $\delta \vB$).}
    %
    \begin{equation}
        \vec{F}_{\rm m \rightarrow v }^{(EM)} = 
         m_p \varkappa  \int\limits_R d^2\rho \,  \left[\vec{e}_z \times  \left( \delta \vj_e  -  \delta \vj_p \right)  \frac{\mathcal{P}'(\rho)}{2 \pi \rho}
         - \frac{e}{c} \frac{n_{p0}}{m_p} \delta \vB \times \vec{e}_\phi  \,  \frac{ 1 - \mathcal{P}(\rho)}{2 \pi \rho}
         \right].
    \end{equation}
    Taking into account the screening condition \eqref{eq:screen_cond_npe}, one can verify that
    $\vec{F}_{\rm{LM}p} + \vec{F}_{\rm{LE}p} + \vec{F}_{\rm{LM}e} + \vec{F}_{\rm{LE}e} +  \vec{F}_{\rm m \rightarrow v }^{(EM)} =0$.
    Hence, the total force on the vortex is equal to the Magnus force \eqref{eq:FMp} at any distance from the vortex axis.

\section{Backflow contribution to the proton current}
\label{sec:backflow}

    In Sec.~\eqref{sec:protons}, the correction to the proton current density caused by the electron scattering off the vortex was determined. 
    In this appendix, we briefly discuss corrections that can be added to that solution.
    Since the electron current density correction $\delta \vj_e$ has already been accounted by solution \eqref{eq:djp_reg_def}, our additional correction should satisfy the following homogeneous set of equations:
    \begin{align}
        \label{eq:RotB_ap}    
        &\Rot \delta \vB_{\rm bf} = \frac{4\pi e}{c} \delta \vj_{p,\rm bf},
        \\
        \label{eq:Rotj_ap}        
        &\Rot \delta \vj_{p,\rm bf} = - \frac{e}{c}\frac{n_{p0}}{m_p} \delta \vB_{\rm bf},
        \\
        \label{eq:DivB_ap}
        &\Div \delta \vB_{\rm bf} = 0,
        \\
        \label{eq:Divj_ap}
        &\Div \delta \vj_{p,\rm bf} = 0.
    \end{align}
    Here, the subscript ``${\rm bf}$'' stands for ``backflow''.
    This term highlights the analogy between the solution of the homogeneous equations \eqref{eq:RotB_ap}--\eqref{eq:Divj_ap} and the backflow known in conventional hydrodynamics. 
    The latter phenomenon occurs near the surface of a body streamlined by the fluid, ensuring the satisfaction of the boundary conditions on the body's surface.
    Following the analogy with this phenomenon,
    the quantities $\delta \vB_{\rm bf}$ and $\delta \vj_{p,\rm bf}$, as in Sec.\eqref{sec:protons}, are assumed to be linear in the incident particle fluxes. 
    This allows us to substitute the unperturbed proton density $n_{p0}$ in Eq.\ \eqref{eq:Rotj_ap}.
    Generally, for our problem the current density and magnetic field corrections can be represented as
    \begin{align}
        \label{eq:djp_rep_bf}
        &\delta \vj_{p,\rm bf} = - \vec{e}_z \times \nabla \Phi +  \Psi \vec{e}_z,
        \\
        \label{eq:dB_rep_bf}
        &\delta \vB_{\rm bf} = \frac{4\pi e}{c}\, \Phi \, \vec{e}_z
            + \frac{c}{e} \frac{m_p}{n_{p0}} \vec{e}_z \times \nabla \Psi,
    \end{align}
    where $\Phi$ and $\Psi$ are some functions of the coordinates $\rho$ and $\phi$ to be determined.  
    Indeed, it is easy to see, that Eqs.~\eqref{eq:DivB_ap} and \eqref{eq:Divj_ap} are automatically satisfied, while Eqs.~\eqref{eq:RotB_ap} and \eqref{eq:Rotj_ap}  bring us to two identical equations for the functions $\Phi$ and $\Psi$:
    \begin{align}
        &\left(\Delta - \lambda^{-2} \right) \Phi = 0,
        \\
        &\left(\Delta - \lambda^{-2} \right) \Psi = 0.     
    \end{align}
    The solutions to these equations, which vanish at infinity, can generally be represented as
    \begin{align}
        \label{eq:Phi_bf}
        \Phi = \sum_{m=0}^{\infty} \Phi_m  \frac{\sin\left[m(\phi - \phi_m)\right]}{\lambda^m} K_m\left( \frac{\rho}{\lambda}\right),
        \\
        \label{eq:Psi_bf}
        \Psi = \sum_{m=0}^{\infty} \Psi_m  \frac{\sin\left[m(\phi - \psi_m)\right]}{\lambda^m} K_m\left( \frac{\rho}{\lambda}\right).    
    \end{align}
    Here, $K_m$ are the modified Bessel functions of the second kind, while $\Phi_m$, $\Psi_m$, $\phi_m$, and $\psi_m$ are constants that cannot be determined within the framework of our model. 
    Note that the $m=0$ harmonic in the function $\Phi$ describes a purely azimuthal flow, which is already represented by the quantities $\vQ_{pv}$, $\vj_{pv}$, and $\vB_v$.
    Since the azimuthal flow is quantized, a small correction to $\vj_{pv}$ cannot exist.
    Thus, the coefficient $\Phi_0$ must be zero.

    To evaluate the contribution of these new corrections to the force, we utilize the results of Appendix~\ref{sec:fin_dist_force}.
    Specifically, we need to calculate the contributions of protons, electrons, and the electromagnetic field, given by Eqs.~\eqref{eq:ap:Fmvp2}, \eqref{eq:ap:Fmve}, and \eqref{eq:FEM1}, respectively.
    First, we note that the new correction can, in principle, modify the electric field.
    However, as before, the terms describing the interaction with the electric field in Eqs.~\eqref{eq:ap:Fmvp2} and \eqref{eq:ap:Fmve} cancel each other.
    Second, the correction to the current determined by the function $\Psi$ is directed along the $z$-axis while the corresponding correction to the magnetic field lies in the perpendicular plane.
    These corrections produce zero force being substituted into Eqs.~\eqref{eq:ap:Fmvp2} and \eqref{eq:FEM1}.
    Note that these equations account only for the momentum flux through the lateral surface of the cylinder.
    However, since the solution does not depend on $z$-coordinate the sum of momentum fluxes through the top and bottom surfaces of the cylinder is always zero as well.

    Substituting representation \eqref{eq:djp_rep_bf} with 
    $\Phi$ given by 
    \eqref{eq:Phi_bf} into the first term of Eq.~\eqref{eq:ap:Fmvp2}, we observe that only the dipolar ($m=1$) harmonic remains after integration over the azimuthal angle $\phi$, yielding:
    \begin{align}
        \nonumber
        \vec{F}_{\rm m \rightarrow v }^{(p)}  &= -  {m_p \varkappa}
        {\Phi_1} \int\limits_0^{2\pi} \frac{d\phi}{2\pi} \, \nabla  \left[ \frac{\sin(\phi - \phi_1)}{\lambda} K_1\left(\frac{R}{\lambda}\right) \right] \frac{R}{\lambda} K_1 \left(\frac{R}{\lambda}\right)
        \\
       & =    {m_p \varkappa}
       \frac{\Phi_1}{2\lambda^2} K_0\left(\frac{R} {\lambda}\right)\frac{R}{\lambda} K_1 \left(\frac{R}{\lambda}\right) \vec{e}_z \times \vec{e}_1,
       \label{eq:Fnvp_bf2}
    \end{align}
    where $\vec{e}_1$ is the unit vector defined by the angle $\phi_1$ and we have substituted Eq.~\eqref{eq:P_apr}.
    The contribution from electrons reduces to its electric part, Eq.~\eqref{eq:FLEe}.
    As already mentioned, it cancels out with the corresponding term in the proton contribution.
    The third component is the force produced by the electromagnetic field.
    By substituting representation~\eqref{eq:dB_rep_bf} with function~\eqref{eq:Phi_bf} into Eq.~\eqref{eq:FEM1}, we observe once again that all harmonics average out to zero, except for the dipolar one.
    Thus, the electromagnetic contribution can be expressed as
    \begin{equation}
         \vec{F}_{\rm m \rightarrow v }^{(EM)} = - {m_p \varkappa}  \frac{R}{4\pi} \frac{1}{2\pi\lambda^2} K_0\left(\frac{R}{\lambda}\right) \int\limits_0^{2\pi} \frac{d\phi}{2\pi} \vec{e}_\rho \,  \Phi_1 \frac{4\pi e}{c}  
         \frac{\sin(\phi - \phi_1)}{\lambda} K_1\left(\frac{R}{\lambda}\right),
    \end{equation}
    where we made use of Eq.~\eqref{eq:dP_apr}.
    By performing the integration over the azimuthal angle, one can easily verify
     that this force is exactly equal to Eq.~\eqref{eq:Fnvp_bf2}, but with the opposite sign.
    Thus, we can conclude that, within the approximation linear in the incident fluxes, these additional corrections do not affect the force.

    Finally, let us consider a formal limit of $\lambda \rightarrow \infty$ corresponding to the case of an uncharged superfluid. 
    Using the well-known asymptotes of Bessel functions $K_m(x)$ at $x\rightarrow 0$, 
    we obtain the following expression for the current correction given by the function $\Phi$:
    \begin{align}
        \nonumber
        \delta \vj_{p, \rm bf}
        &= 
        -  
        \vec{e}_z \times \nabla \sum_{m=1}^{\infty} \Phi_m  {2^{m-1}(m-1)!} \frac{\sin\left[m(\phi - \phi_m)\right]}{{\rho^m}} 
        \\
        & =  
        \nabla \sum_{m=1}^{\infty} \Phi_m  {2^{m-1}(m-1)!} \frac{\cos\left[m(\phi - \phi_m)\right]}{{\rho^m}}.
        \label{eq:_bf_uncharged} 
    \end{align}
    We find  that it takes the form of a potential flow similar to the backflow known in conventional hydrodynamics.
    For a streamlined body with a cylindrical shape, the backflow reduces to the dipolar harmonic \cite{SoninBook}.
    It is interesting to note that in the charged superfluid, the backflow (if present) acquires a length-scale given by the London penetration depth.

\section{Accounting for the entrainment effect}
\label{sec:entr}

    It is generally accepted that a phenomenon known as entrainment effect can occur in a mixture of two (or more) superfluids \cite{AndreevBashkin1976}.
    For $npe$-matter this effect manifests in the fact that each neutron and proton particle current densities, $\vj_n$ and $\vj_p$,
    can generally depend on both condensate momenta, $\vQ_n$ and $\vQ_p$. 
    In the limit of vanishing temperature this implies
    \begin{align}
    \label{eq:jn_def_entr}
        \vj_n & = Y_{np} \vQ_p + Y_{nn} \vQ_n,
        \\
        \label{eq:jp_def_entr}
        \vj_p & = Y_{pp} \vQ_p + Y_{pn} \vQ_n, 
    \end{align}
    where 
    $Y_{\alpha\alpha'}$ is a symmetric  matrix.
    Note that, instead of the standard nonrelativistic entrainment (or Andreev-Bashkin)  matrix $\rho_{\alpha\alpha'}$ \cite{AndreevBashkin1976}, we employ here the matrix $Y_{\alpha\alpha'}$, which was initially introduced and utilized  in the context of relativistic superfluid hydrodynamics (see, e.g., Refs.\ \cite{GusakovKantorHaensel2009,GusakovKantorHaensel2009b,kg11, GusakovDommes2016}).
    In the non-relativistic limit, these two matrices are related by the expression: 
    \begin{equation}
        \rho_{\alpha\alpha'} = m_\alpha m_{\alpha'}  Y_{\alpha\alpha'}.
    \end{equation}
    At vanishing temperature the matrix $Y_{\alpha\alpha'}$ satisfies a sum rule (see, e.g., Ref.\ \cite{GusakovKantorHaensel2009}) which, in the nonrelativistic limit, can be represented as
    \begin{equation}
        \label{eq:sum_rule}
        \sum_{\alpha'} m_{\alpha'} Y_{\alpha\alpha'} = n_\alpha,
    \end{equation}
    and is equivalent to the well-known sum rule for the Andreev-Bashkin matrix, $\sum_k \rho_{ik}=\rho_i$.
    In this appendix, we briefly discuss how the entrainment effect 
    affects our derivation of the force acting on a vortex.

    First, the representation of the proton condensate momentum, given by Eq.\ \eqref{eq:Q_mod_ch}, remains valid. 
    In contrast, the neutron condensate momentum is no longer reduced to the homogeneous incident flux but should contain a correction:
    \begin{equation}
        \vQ_n = \vQ_{n0} + \delta \vQ_n.
    \end{equation}   
    Eqs.\ \eqref{eq:Q_def}, \eqref{eq:mu_def}, and, consequently, Eq.\ \eqref{eq:sf_eq1} remain valid. 
    The continuity equations still have the form of Eqs.\ \eqref{eq:cont_eq_p_td} and \eqref{eq:cont_eq_n_td}, but with the current densities given by Eqs.\ \eqref{eq:jn_def_entr} and \eqref{eq:jp_def_entr}.

    The Euler-like form of the superfluid equation, Eq.\ \eqref{eq:sf_eq2}, should, in turn, be replaced by a slightly more complex equation:
    \begin{equation}
        \label{eq:sf_eq2_entr}
        \frac{\partial \vQ_p}{\partial t} + \frac{1}{m_p} ( \vQ_p \nabla ) \vQ_p 
        -  \frac{m_p m_n Y_{np}}{n_p}  \frac{\Rot\vQ_p}{m_p} \times\left( \frac{\vQ_p}{m_p} - \frac{\vQ_n}{m_n} \right)
        +\nabla \left( \breve{\mu}_p - \frac{\vQ_p^2}{2 m_p}\right)
         = \vec{f}_{\rm Lp} - \vec{f}_{\rm M},
    \end{equation}  
    where the forces on the right-hand side are given by Eqs.\ \eqref{eq:Lorenz_def} and \eqref{eq:Magnus_per_p}, but again with the proton current density expressed in Eq.\ \eqref{eq:jp_def_entr}.
    Equation \eqref{eq:sf_eq2_entr} can be derived by adding and subtracting the term
    \begin{equation}
         \frac{m_p m_n Y_{np}}{n_p}  \frac{\Rot\vQ_p}{m_p} \times\left( \frac{\vQ_p}{m_p} - \frac{\vQ_n}{m_n} \right)
    \end{equation}
    to/from Eq.\ \eqref{eq:sf_eq1}, taking into account Eq.\ \eqref{eq:sum_rule} and performing algebraic transformations similar to those used in deriving Eq.\ \eqref{eq:sf_eq2}.
    The neutron counterpart of Eq.\ \eqref{eq:sf_eq2_entr} can be obtained from this equation by exchanging $n \leftrightarrow p$ and setting $e=0$:
    \begin{equation}
        \label{eq:sf_eq2n_entr}
        \frac{\partial \vQ_n}{\partial t} + \frac{1}{m_n} ( \vQ_n \nabla ) \vQ_n 
        +  \frac{m_p m_n Y_{np}}{n_p}  \frac{\Rot\vQ_n}{m_n} \times\left( \frac{\vQ_p}{m_p} - \frac{\vQ_n}{m_n} \right)
        +\nabla \left( \breve{\mu}_n - \frac{\vQ_n^2}{2 m_n}\right)
         = 0.
    \end{equation} 
    In fact, $\Rot\vQ_n$ is zero. However, we retain this formal term to simplify the derivation of the momentum conservation equation.
    A set of equations similar to Eqs.\ \eqref{eq:sf_eq2_entr} and \eqref{eq:sf_eq2n_entr} can be found, e.g., in Ref.\ \cite{Mendell1991a}.%
    %
    \footnote{Note that, in this reference, a phenomenological coarse-grained set of equations is derived. Thus, the functions $\Rot \vQ_\alpha$ are treated as continuous.
    The chemical potentials should also be redefined as $\tilde{\mu}_\alpha = \breve{\mu}_\alpha - {\vQ_\alpha^2}/{2 m_\alpha}$, where $\tilde{\mu}_\alpha$ are the chemical potentials from Ref.\ \cite{Mendell1991a}. }
    %

    The momentum conservation equation can be derived in the way described in Appendix \ref{sec:mom_cons}.
    It has the form provided by Eq.\ \eqref{eq:mom_eq_ap}, where $\Pi_{e}^{ik}$ and  $\Pi_{EM}^{ik}$ are given by Eqs.\ \eqref{eq:Pi_e_ap} and \eqref{eq:Pi_EM_ap}, respectively, whereas 
    \begin{equation}
        \label{eq:Pi_np_entr}
        \Pi_{np}^{ik} = \sum_{\alpha\alpha'} Y_{\alpha\alpha'} Q_{\alpha}^i Q_{\alpha'}^k  + \delta^{ik} P_{np}.
    \end{equation}
    In the last expression, we defined the neutron-proton pressure, whose differential 
    equals
    \begin{equation}
        \label{eq:dPnp_entr}
        d P_{np} = \sum_{\alpha} n_\alpha d \breve{\mu}_\alpha - \sum_{\alpha} \vj_\alpha d \vQ_\alpha.
    \end{equation}
    An expression for the stress tensor, analogous to Eq.\ \eqref{eq:Pi_np_entr}, can be found in the work of Andreev and Bashkin \cite{AndreevBashkin1976}, where the normal density should be set to zero.
    In that paper, the chemical potentials are defined in a reference frame moving with the normal component.
    However, since we are working in the limit of vanishing temperature, there is technically no normal component in our problem.
    Nonetheless, if we formally introduce an abstract normal component moving with velocity $\vec{u}$, the Andreev-Bashkin chemical potentials, $\mu_{\alpha}^{\rm AB}$, can be related to ours through the following equation:
    \begin{equation}
        \breve{\mu}_\alpha = m_\alpha \mu_{\alpha}^{\rm AB} + \vQ_\alpha \vec{u} - \frac{m_\alpha u^2}{2}.
    \end{equation}    
    With this replacement, the pressure differential \eqref{eq:dPnp_entr} clearly coincides with the corresponding expression in the Andreev-Bashkin theory.
    Alternatively (and more straightforwardly), one can assume that this abstract normal component is at rest in the vortex frame (${\pmb {\rm u}}=0$).
    In this case, $\breve{\mu}_\alpha = m_\alpha \mu_{\alpha}^{\rm AB}$, and Eq.\ \eqref{eq:dPnp_entr} immediately takes the form of the corresponding equation from Andreev and Bashkin's work \cite{AndreevBashkin1976}.

    The proton current density, with the accuracy adopted in this paper, can be expressed in the form given by Eq.\ \eqref{eq:jp_rep},
    where now
    \begin{equation}
        \label{eq:jp0_entr}
        \vj_{p0} = Y_{pp,0} \vQ_{p0} + Y_{pn,0} \vQ_{n0} 
    \end{equation}
    is the incident transport current,
    \begin{equation}
        \label{eq:jpv_entr}
        \vj_{pv} = Y_{pp,0} \vQ_{pv} 
    \end{equation}
    is the current generated by the vortex with $\vQ_{pv}$ given by \eqref{eq:Q_pv_def}, and, finally, 
    \begin{equation}
        \label{eq:djp_entr}
        \delta \vj_{p} = Y_{pp,0} \delta \vQ_{p} +  Y_{pn,0} \delta \vQ_{n} 
    \end{equation}
    is the self-consistent correction. 
    In these expressions, the coefficients $Y_{\alpha\alpha',0}$ stand for the unperturbed entrainment matrix. 
    The incident flux satisfies the screening condition \eqref{eq:screen_cond_npe}.

    Reproducing the calculations from Appendix \eqref{sec:mag_field} with $\vj_{pv}$ given by Eq.\ \eqref{eq:jpv_entr}, one can verify that the expression for the vortex magnetic field remains the same, except for the London penetration depth, which is now given by \cite{Mendell1991a,GusakovDommes2016}
    \begin{equation}
	\label{eq:lambda_def_entr}
	\lambda = \sqrt{\frac{c^2}{4\pi e^2 Y_{pp,0}}}.
    \end{equation}
    Thus, the problem of electron scattering off the vortex magnetic field remains unaffected.
    As for the protons, it can be verified that all the equations from Sec.\ \ref{sec:protons} formally remain valid if one replaces the definitions of the current density components \eqref{eq:jp0_def}--\eqref{eq:djp_def} with Eqs.\ \eqref{eq:jp0_entr}--\eqref{eq:djp_entr}.
    The screening condition, Eq.\ \eqref{eq:screen_cond_npe}, is also unchanged.
    Hence, the correction $\delta \vj_{p} $ is still given by Eq.\ \eqref{eq:djp_atline}.

    Since the only impact of the entrainment effect on electron scattering is the redefinition of the London penetration depth, the electron contribution to the force calculated with Method I (Sec.\ \ref{sec:Method_I}) is unaltered, apart from corrections to $\lambda$.
    Furthermore, an inspection of the nuclear stress tensor \eqref{eq:Pi_np_entr} reveals that, as in the case without entrainment, the only neutron-proton contribution arises from the pressure and corresponds to Eq.\ \eqref{eq:Fmv_np}.
    The reasoning that justifies neglecting the electromagnetic contribution also holds true.
    Thus, we can conclude that the impact of the entrainment effect reduces to the redefinition of the London penetration depth in the longitudinal force [see Eqs.\ \eqref{eq:F_par_meth_I} and \eqref{eq:Le_val}].
    The same result can be obtained with Method II.
    Indeed, the Magnus force depends on proton current density, while all the expressions for the proton current density components remain formally intact, as we have just discussed.

\bibliography{mn-jour,paper}

\end{document}